\documentclass[11pt,a4paper]{article}
\pdfoutput=1
\usepackage{jheppub}
\usepackage{gensymb}
\usepackage{multirow}
\usepackage{tabularx}
\usepackage[table,xcdraw]{xcolor}
\DeclareSymbolFont{matha}{OML}{txmi}{m}{it}% txfonts
\DeclareMathSymbol{\varv}{\mathord}{matha}{118}
\usepackage{amsmath}
\usepackage{amssymb}
\usepackage{graphicx}
\usepackage{subfigure}
\usepackage{pstricks}
\usepackage{bm}
\usepackage{pbox}
\usepackage{placeins}
\usepackage{booktabs}
\usepackage[T1]{fontenc}
\usepackage{footnote}
\usepackage{pdfpages}
\usepackage{hhline}
\usepackage{multirow}
\usepackage{multicol}
\usepackage{enumitem}
\usepackage[toc,page]{appendix}
\allowdisplaybreaks
\usepackage{comment}
\usepackage{float}                                             
\usepackage{mathtools}                  
\usepackage{textcomp}
\usepackage{gensymb}

\usepackage{relsize}

\usepackage{slashed}

\usepackage{graphicx}
\usepackage{float}
\usepackage{xcolor}
\usepackage[utf8x]{inputenc} 
\usepackage[normalem]{ulem}

\usepackage[numbers]{natbib}
\usepackage{notoccite}

\usepackage{rotating}
\usepackage{tabularx}
\usepackage[labelfont=bf]{caption} % optional

\FloatBarrier
\usepackage[many]{tcolorbox}

\renewcommand{\thetable}{\Roman{table}}

\definecolor{MyDarkBlue}{rgb}{0.1, 0.1, 0.8} %defining the color 'MyDarkBlue'
\definecolor{SBlue}{rgb}{0.2, 0.4, 0.7} %defining the color 'MyDarkBlue'
\definecolor{MyLightBlue}{rgb}{0.22,0.51,0.9}
\definecolor{MyGreen}{rgb}{0.0, 0.5, 0.0}
\definecolor{BrickRed}{rgb}{0.8, 0.25, 0.33}
\RequirePackage{hyperref}
\hypersetup{colorlinks, citecolor=MyGreen,linkcolor=MyDarkBlue, urlcolor=BrickRed}

\newcommand{\lhs}{\lambda_\text{HS}}
\newcommand{\ls}{\lambda_S}
\newcommand{\lh}{\lambda_H}

\usepackage{tcolorbox}
\begin{document}
%=====================================================
\title{
FIMP Dark Matter from Flavon Portals
 }
  \author[a]{K.S. Babu \footnote{babu@okstate.edu},}
 \author[b]{~~Shreyashi Chakdar\footnote{schakdar@holycross.edu},}
 \author[c]{~~Nandini Das\footnote{nandinidas.rs@gmail.com},}
 \author[c]{~~Dilip Kumar Ghosh\footnote{dilipghoshjal@gmail.com},}
 \author[ c]{~~~~Purusottam Ghosh\footnote{spspg2655@iacs.res.in},}

 \affiliation[a]{Department of Physics, Oklahoma State University, Stillwater, OK 74078, USA}

 \affiliation[b]{Department  of Physics, College of the Holy Cross, Worcester, MA 01610, USA}
 
\affiliation[c]{School of Physical Sciences, Indian Association for the Cultivation of Science,\\ 2A \& 2B Raja S.C. Mullick Road, Kolkata, 700032, India}

%\preprint{OSU-HEP-20-15}

%%%%%%%%%%%%%%%%%%%%%%%%%%%%%%%%%%%%%%%%%%%%%%%%%%%%%%%%%%%%%%%%%%%%%%%%%%%%%%%%%%%%%

\abstract{We investigate the phenomenology of a non-thermal dark matter (DM) candidate in the context of flavor models that explain the hierarchy in the masses and mixings of quarks and leptons via the Froggatt-Nielsen (FN) mechanism.  A flavor-dependent $U(1)_{\rm FN}$ symmetry explains the fermion mass and mixing hierarchy, and also provides a mechanism for suppressed interactions of the DM, assumed to be a Majorana fermion, with the Standard Model (SM) particles, resulting in its FIMP (feebly interacting massive particle) character. Such feeble interactions are mediated by a flavon field  through higher dimensional operators governed by the $U(1)_{\rm FN}$ charges. We point out a natural stabilizing mechanism for the DM within this framework with the choice of half-integer $U(1)_{\rm FN}$ charge $n$ for the DM fermion, along with integer charges for the SM fermions and the flavon field. In this flavon portal scenario, the DM is non-thermally produced from the decay of the flavon in the early universe which becomes a relic through the freeze-in mechanism. We explore the allowed parameter space for this DM candidate from relic abundance by solving the relevant Boltzmann equations.
We find that reproducing  the correct relic density requires the DM mass to be in the range $(100-300)$ keV for $n=7.5$ and $(3-10)$ MeV for $n=8.5$ where $n$ is the $U(1)_{\rm FN}$ charge of the DM fermion.
}
%%%%%%%%%%%%%%%%%%%%%%%%%%%%%%%%%%%%%%%%%%%%%%%%%%%%%%%%%%%%%%%%%%%%%%%%%%%%%%%%%%%%%%%%%

\maketitle
%\newpage
%%%%%%%%%%%%%%%%%%%%%%%%%%%%%%%%%%%%%%%%%%%%%%%%%%%%%%%%%%%%%%%%%%%%%%%%%%%%%%%%%%%%%%%%%%%%%%%%%%%%%%%%%%%%%%%%%%%%%%%%%%%%%%%%%%%%%%%%%%%%%%%%%%%%%%%

%%%%%%%%%%%
\section{Introduction}
\label{sec:intro}
The Standard Model (SM) of particle physics is a remarkably successful theory of the modern scientific era. The discovery of the  Higgs boson with a mass $m_h \sim 125$ GeV at the LHC in 2012 \cite{Chatrchyan:2012xdj,Aad:2012tfa} with its properties aligned with the predictions of the SM has strongly validated the theory. However, a few fundamental problems remain unexplained within the SM, which are compelling motivations to consider its extensions.  A prime example is a lack of understanding of the pattern of quark and lepton masses and mixings.  The strong hierarchy spanning some six orders of magnitudes in the charged fermion masses, as well as the disparate mixing angles in the quark and lepton sectors are unexplained within the SM.  Many aspects of the flavor puzzle may be understood 
via the Froggatt-Nielson (FN) mechanism \cite{froggatt1979hierarchy}
which extends the SM to include a flavor $U(1)_{\rm FN}$ symmetry which may be global or local.  Effective Yukawa interactions are induced in powers of a complex scalar field $S$, called the flavon, which is a singlet of the SM gauge symmetry, but is charged  under the $U(1)_{\rm FN}$ flavor symmetry.  The fermions of the SM transform under this $U(1)_{\rm FN}$ with flavor-dependent charges. The Yukawa interactions arise in this framework as effective non-renormalizable operators with powers of the flavon field suppressed by appropriate powers of a cut-off scale $\Lambda$. Once the flavon field acquires a nonzero vacuum expectation value (VEV), fermion masses and mixings are generated with flavor-dependent suppression factors that go as powers of a small expansion parameter $\epsilon \equiv \langle S \rangle/\sqrt{2}\Lambda$, thereby explaining the hierarchy in the masses and mixings. Considerable efforts have gone into implementing this mechanism in beyond the Standard Model scenarios to explain the intricate patterns of quark and lepton masses and mixings (for reviews see  \cite{Babu:2009fd,Feruglio:2015jfa,Xing:2020ijf}). 

The nature of the dark matter in the universe is another profound question that is not explained within the Standard Model. Astrophysical and cosmological observations, such as the rotation curves of galaxies in galaxy clusters, bullet cluster, gravitational lensing and  anisotropy of the cosmic microwave background radiation (CMBR)  \cite{Rubin:1970zza,Spergel:2006hy,Hu:2001bc,Bertone:2004pz,Planck:2018vyg,Roszkowski:2017nbc} have provided strong evidence for the existence of DM in the present universe.  Satellite borne experiments WMAP \cite{Spergel:2006hy} and PLANCK have precisely measured the number density of dark matter to be $\Omega_{\rm DM} h^2 = 0.120 \pm 0.001$ \cite{Planck:2018vyg}, which provides roughly a quarter of the energy budget of the universe.   But the nature of the DM (specifically its spin, mass and its interactions with ordinary matter) is still a mystery. So far, the existence of DM has been inferred directly from gravitational interactions only. 

If the DM has any interaction other than gravity, its strength should be similar to or weaker than that of weak interactions. DM particles are broadly classified into WIMP -- weakly interacting massive particle -- (for review see Ref. \cite{Bertone:2004pz}), SIMP (strongly interacting massive particle) \cite{Dimopoulos:1989hk,Hochberg:2014dra} and FIMP (feebly interacting massive particle) \cite{Hall:2009bx} based on its interaction strengths and production mechanisms.  The WIMP scenario has been very popular due to its detection possibilities in the direct, indirect and collider search experiments. 
However, the null results from dark matter direct search experiments such as LUX \cite{LUX:2016ggv}, XENON \cite{XENON:2018voc} and PANDAX \cite{PandaX-II:2016vec} have set stringent bounds over a wide range of WIMP masses and couplings. These results have provided impetus to explore other viable scenarios for the DM which are less sensitive to direct search experiments. One interesting direction is to look for light dark matter with tiny couplings to the particles in the thermal bath which never attained thermal equilibrium in the early universe, but were produced at a later time from the annihilation and decays of particles in the thermal bath. Here, unlike the WIMP scenario, the dark matter freezes-in to give the correct relic density. This class of particles which typically have lighter masses compared to the WIMP, constitute FIMP \cite{Hall:2009bx}, which is the primary focus of this paper.

In this paper we propose a unified solution to the fermion mass hierarchy puzzle and the (FIMP) dark matter puzzle within a class of $U(1)_{\rm FN}$ extensions of the SM.  The $U(1)_{\rm FN}$ symmetry serves dual purpose here: it explains the pattern of quark and lepton masses and mixings via the Froggatt-Nielsen mechanism, and it also provides a natural scenario for a feebly interacting DM candidate.  The latter arises since the DM candidate $\chi$, which is assumed to be a Majorana fermion in this work, carries a $U(1)_{\rm FN}$ charge, and acquires a suppressed mass (compared to a more fundamental scale $\Lambda$) by the same dynamics as the FN mechanism.  The flavon filed $S$ connects $\chi$ with the SM sector, but the Yukawa coupling of $S$ with $\chi$ is suppressed by a small parameter $\epsilon^{2n}$ where $n$ is the $U(1)_{\rm FN}$ charge of $\chi$ (in a normalization where the FN-charge of $S$ is $-1$).  This naturally leads to feeble interactions of the dark matter particle with the SM particles. It is also possible that the DM is thermal in this framework, corresponding to smaller values of the charge $n$ of $\chi$ ($n\sim 1-3$), but here we focus on the FIMP scenario corresponding to larger values of $n \sim 7-9$.  In this case it is natural for the DM to be relatively light, owing to the FN suppression. We shall see that a DM mass in the range of a few keV to a few MeV is also preferred from its relic abundance. Owing to its feeble couplings $\chi$ never achieved thermal equilibrium in the early universe, but it was produced at late times from the decays (or scatterings) of some heavy particles in the thermal bath. We also show that by choosing the charge $n$ of $\chi$ to be a half-integer, while  the SM fields and the flavon $S$ have integer charges, the DM becomes stable, thus explaining an important question for any DM model. It is worth mentioning here that there are different kind of flavor DM scenarios present in literature in the name of {\it{flavored dark matter}} (see in Ref. \cite{Agrawal:2011ze,Kile:2013ola,Bishara:2015mha}) where DM particles are associated with flavor charges.   

Froggatt-Nielsen mechanism has been applied in the context of thermal WIMP dark matter scenario in Ref. \cite{Calibbi:2015sfa,Alvarado:2017bax} , but has received less attention in connection with the FIMP scenario that we develop here. While this paper was being prepared Ref. \cite{Cheek:2022yof} appeared  which has discussed FIMP dark matter in the context of flavon. There it is assumed that the flavon is produced {\it non-thermally  via freeze-in} and through field oscillations. The fermion dark matter is also produced non-thermally and sequentially from the flavon.  In contrast, in the present work, the flavon is in thermal equilibrium with the SM particles via the dimension-six Yukawa interaction as well as Higgs portal interactions present in these models.  The DM is produced in our framework non-thermally from the thermal flavon via the standard freeze-in mechanism. The resulting parameter space is complementary to those of Ref. \cite{Cheek:2022yof}.
FIMP dark matter has been studied in a general context in various models, unrelated to the fermion mass and mixing hierarchies, see for e.g. Ref. \cite{Bernal:2017kxu,Molinaro:2014lfa,Kang:2014cia,Hessler:2016kwm,Biswas:2016yjr,DEramo:2020gpr,Liu:2022jdq}.

The paper is organised as follows.  In Sec. \ref{sec:model} we discuss the model framework. Here we present two scenarios for the $U(1)_{\rm FN}$ charges and show how the dark matter fermion $\chi$ can be incorporated.  We also discuss here the flavor violation induced by the flavon field and show its consistency. In Sec. \ref{sec:DMphen} we develop the phenomenology of the FIMP dark matter.  Here we summarize theoretical and experimental constraints on the model parameters and discuss the freeze-in  mechanism of $\chi$ in the early universe.  Our numerical results for the relic abundance of $\chi$ are presented here, obtained from the solutions to the Boltzmann equations, assuming that the scalar and pseudoscalar flavon portals contribute equally. In Sec. \ref{sec:pseudo} we focus on the case where only the pseudoscalar flavon is active and analyze the DM relic density. In Sec. \ref{sec:FSL} we discuss the free streaming length of the dark matter and how it can constrain our model parameter space. Finally in Sec.  \ref{sec:concl} we present our conclusions. In several appendices, we provide various technical details on the interaction Lagrangian (\ref{apx:int}), decay widths (\ref{apx:dw}), and the interaction rates (\ref{apx:IntRate}).
 
%%%%%%%%%%%
%%%%%%%%%%%
\section{The Model Framework}
\label{sec:model}
%%%%%%%%%%%

In the Froggatt-Nielsen framework that we study, all Standard Model fermions, except the top quark, are charged under the $U(1)_{\rm FN}$ symmetry (the top quark acquires its mass via a renormalizable coupling, and thus has zero $U(1)_{\rm FN}$ charge).
The dark matter candidate $\chi$, taken to be a SM singlet Majorana fermion, also carries a nonzero $U(1)_{\rm FN}$ charge. The $U(1)_{\rm FN}$ symmetry may be global or local.  If it is local, there are severe constraints on the fermionic charges from anomaly cancellation conditions.  For the most part in our discussions, we shall assume the $U(1)_{\rm FN}$ symmetry to be global, broken softly by dimension-two terms in the scalar potential.  We also suggest an intriguing scenario where the $U(1)_{\rm FN}$ symmetry acts as a local flavor symmetry, wherein the triangle anomalies cancel automatically,  without the introduction of any exotic fermion. In all cases, a complex scalar field $S$, which is a singlet of the SM gauge symmetry, is introduced  with flavor charge denoted by $a_S$, which plays a crucial role in explaining  the hierarchical masses of fermions and the DM particle. The interaction Lagrangian of the SM particles with the extended symmetry  $SU(2)_L \otimes U(1)_Y \otimes U(1)_{\rm FN} $ can be written, in the case of global $U(1)_{\rm FN}$, as
\begin{equation}
   \mathcal{L}=  \mathcal{L_{\rm Scalar}^{\rm Flavon+Higgs}}+\mathcal{L_{\rm DM}}+\mathcal{L_{\rm SM}}^{Kin}+\mathcal{L_{\rm SM}}^{\rm FN-Yuk}~.
    \end{equation}
We shall analyze each of the new interaction terms in the Lagrangian in turn. 
    \subsection{Scalar Sector} The Lagrangian for the 
    scalar sector of the model which includes the SM Higgs doublet $H$ and a complex scalar singlet field $S$,  the flavon, is given by     
    \begin{equation}
\mathcal{L_{\rm Scalar}^{\rm Flavon+Higgs}} = \left|D_\mu\,H\right|^2 + \left|\partial_\mu\,S\right|^2-V(H,S)\,.
\label{eq:ls}
\end{equation}
The Higgs potential is given by (see for e.g. Ref. \cite{Coito:2021fgo}):
\color{black}
\begin{equation}
V(H,S) = -\mu_H^2\,\left|H\right|^2 + \lh\,\left|H\right|^4 - \mu_S^2\,\left|S\right|^2 + \ls\,\left|S\right|^4 + \lhs\,\left|H\right|^2\,\left|S\right|^2\,-\mu^2_b (S^2+ h.c.)
\label{eq:pot}
\end{equation}
Here we have included a soft $U(1)_{\rm FN}$ breaking term $\mu_b^2$ in the absence of which a Goldstone boson would result upon the spontaneous breaking of the $U(1)_{\rm FN}$.  While terms such as $(S^3 + h.c.)$,  ($\mu^3 S + h.c.)$ and $(|H|^2 S + h.c.)$ are also soft breaking terms, we do not include them in our study, since omitting these terms which are odd under $S \rightarrow -S$
reflection symmetry (which may be extended to the full Lagrangian) is self-consistent with renormalizability.   

For $\mu_H^2 > 0$ and $\mu_S^2> 0$ in Eq. (\ref{eq:pot}), we can parametrize the fields $S$ and $H$ as
\begin{eqnarray}
    S=\frac{1}{\sqrt{2}}(h_S+v_S+i A_S),\,~~~
H=
\frac{1}{\sqrt{2}}\begin{pmatrix}
0 \\ (h_H +v_H)
\end{pmatrix}\,
\end{eqnarray}
with $v_H \simeq 246$ GeV.
Minimizing the potential leads to the conditions (in the spontaneously broken symmetric phase)
\begin{align}
&\mu_H^2 = v_H^2\,\lambda_H+\frac{1}{2}\,\lambda_{HS}\,v_S^2\,
\nonumber\\&
\mu_S^2 = v_S^2\,\lambda_S + \frac{1}{2}\,\lambda_{HS}\,v_H^2\, -2 \mu^2_b~. 
\end{align}
The mass of the CP odd scalar field $A_S$, which has no mixing with the CP even scalar fields, is proportional to the soft $U(1)_{\rm FN}$ breaking term $\mu_b^2$, and is given by
\begin{equation*}
  m^2_{A_S} = 4 \mu^2_b .
\end{equation*}
The two CP even states $h_H$ and $h_S$ mix with a mass matrix given by 
\begin{equation}
\mathcal{M}^2=\left(
\begin{array}{cc}
 2 \lambda_H \,v_H^2 & \lambda_{HS}\,v_H\,v_S\\
\lambda_{HS}\, v_H\,v_S & 2 \lambda_S\,v_S^2\\
\end{array}
\right)\,~.
\end{equation}
The eigenvalues of the mass eigenstates $h_1$ and $h_2$ are then
\begin{equation}
m^2_{h_1,h_2} = \big(\lambda_H v^2_H+\lambda_S v^2_S\big) \mp \sqrt{\lambda^2_H v^4_H+\lambda^2_S v^4_S +(\lambda_{HS}- 2 \lambda_{S}\lambda_{H})v^2_H v^2_S} ~.
\end{equation}
These eigenstates are related to the original states via
\begin{equation}
\begin{pmatrix}
h_1 \\ h_2
\end{pmatrix}=
\begin{pmatrix}
\cos\theta & -\sin\theta \\ \sin\theta & \cos\theta
\end{pmatrix}\,
\begin{pmatrix}
h_H \\ h_S
\end{pmatrix}\,,
\end{equation}

\noindent where $h_1$ is identified as the SM-like Higgs with mass $m_{h_1} \simeq 125$ GeV and $h_2$ is a new scalar particle with mass $m_{h_2}$. 
The Higgs mixing angle is given by
\begin{equation}
\tan 2\theta = \frac{\lambda_{HS}\,v_H\,v_S\, }{ \lambda_S \, v_S^2\,-\lambda_H\, v_H^2\, }\,\,. \label{eq:mix}   
\end{equation}
\noindent One may then express the scalar sector parameters of 
$V(H,S)$ in terms of physical masses and the mixing angle $\theta$:
\begin{eqnarray}\label{eq:cplngs}
\ls &=& \frac{1}{2\,v_S^2}\,\left(m_{h_1}^2\,\sin ^2\theta+m_{h_2}^2\,\cos ^2\theta\right)
\nonumber\\
\lh &=& \frac{1}{2\,v_H^2}\,\left(m_{h_1}^2\,\cos ^2\theta+m_{h_2}^2\,\sin^2\theta\right)
\nonumber\\
\lambda_{HS}\, &=& \frac{\sin2\theta \,(m_{h_2}^2-m_{h_1}^2)} {2\,v_H\,v_S} \nonumber \\
\mu_b^2 &=& \frac{1}{4}M_{A_S}^2\,~.
\label{eq:param}
\end{eqnarray}
We note that for small $\sin\theta$, the mass eigenstate $h_2$ is essentially the singlet flavon state $S$. These relations, Eq. (\ref{eq:cplngs}) will be utilized in our DM relic abundance calculations. 

At high temperatures when $T^2 \gg \mu_S^2,\,\mu_b^2$, the theory will be in the $U(1)_{\rm FN}$ symmetric phase where all fermions including the DM particle will be massless (see Sec. \ref{sec:dm}).  At temperatures below the $U(1)_{\rm FN}$ symmetry breaking, but above the electroweak phase transition, all SM fermions will remain massless, but the DM fermion $\chi$ will acquire a mass.  And at temperatures below the electroweak symmetry breaking scale all fermions as well as $W$ and the $Z$ gauge bosons will acquire masses. We shall follow this thermal history in our computation of DM relic abundance.

\subsection{Froggatt-Nielsen Mechanism: The General Framework}

In the FN mechanism, a flavor-dependent $U(1)_{\rm FN}$ symmetry is assumed, under which the SM fermions transform nontrivially. This symmetry is broken by the VEV of a SM singlet flavon field $S$. The Yukawa Lagrangian for the fermions in this framework is given by
\begin{equation}
   \mathcal{L}_{\rm SM}^{\rm FN-Yuk} =  
   -y_{ij}^{(u)}\Big(\frac{S}{\Lambda}\Big)^{n^u_{ij}} Q_i H u^c_j - y_{ij}^{(d)}\Big(\frac{S}{\Lambda}\Big)^{n^d_{ij}} Q_i \tilde{H} d^c_j - y_{ij}^{(e)}\Big(\frac{S}{\Lambda}\Big)^{n^e_{ij}} L_i \tilde{H} e^c_j + h.c.
   \label{eq:FN}
\end{equation}
Here $Q_i$ and $L_i$ are the left-handed quark and lepton doublets, and $(u^c_i,\,d^c_i,\,e^c_i)$ are the left-handed conjugate up-quark, down-quark and charged lepton singlet fields respectively.  ($i,\,j$) here are flavor indices, and $SU(2)_L$ contraction is via $\epsilon_{ab}$ which is not explicitly shown.  The spinor contractions in Eq. (\ref{eq:FN}) are by the charge conjugation matrix $C$, which is also not shown. The parameter $\Lambda$ is an effective cut-off scale where new degrees of freedom arise, including vector-like FN fermion fields.  
$n^u_{ij}$, $n^d_{ij}$ and $n^e_{ij}$ in Eq. (\ref{eq:FN}) are integers related to the  $U(1)_{\rm FN}$ charges of fermions through the relations
\begin{eqnarray}
      n_{ij}^u &=& a_{Q_i} + a_{u_j} + a_H \nonumber \\
      n_{ij}^d &=& a_{Q_i} + a_{d_j} + a_H \nonumber \\
      n_{ij}^e &=& a_{L_i} + a_{e_j} + a_H
      \label{eq:nij}
\end{eqnarray}
where $a_H$ and ($a_{Q_i}$, $a_{L_i}$, $a_{u_i}$, $a_{d_i}$, $a_{e_i}$) are the flavor-dependent $U(1)_{\rm FN}$ charges of the SM Higgs doublet, and the fermion fields $(Q_i,\,L_i,\,u^c_i,\,d^c_i, e^c_i)$ respectively.   In writing Eq. (\ref{eq:nij}) the flavor charge of the flavon field $S$ has been taken to be $a_S = -1$, which can be chosen without loss of generality. We shall also choose $a_H = 0$, which can be achieved by taking a linear combination of hypercharge and the flavor charge. That is, by choosing $a_H' = (a_H - 2\,a_H\, Y_H)$, where $Y_H = 1/2$ is the hypercharge of $H$, the FN-charge of $H$ can be shifted to zero for any integer values of $a_H$.
The dimensionless couplings $y^{(u)}_{ij}$, 
$y^{(d)}_{ij}$ and $y^{(e)}_{ij}$
in Eq. (\ref{eq:FN}) are all taken to be of order unity. The mass matrices for the
up-type quarks, down-type quarks, and charged leptons arise through higher dimensional operators once the $U(1)_{\rm FN}$ symmetry is broken with $\langle S \rangle = v_S/\sqrt{2}$:
\begin{eqnarray}
 m^u_{ij} &=&  y^{(u)}_{ij} \Big(\frac{v_S}{\sqrt{2} \Lambda} 
\Big)^{n^{u}_{ij}}\,\frac{v_H}{\sqrt{2}} \equiv Y^{(u)}_{ij}\, \frac{v_H}{\sqrt{2}}\nonumber \\ 
 m^d_{ij} &=&y^{(d)}_{ij} \Big(\frac{v_S}{\sqrt{2} \Lambda} 
\Big)^{n^{d}_{ij}}\,\frac{v_H}{\sqrt{2}}
 \equiv Y^{(d)}_{ij}\, \frac{v_H}{\sqrt{2}}\nonumber \\
 m^e_{ij} &=&y^{(e)}_{ij} \Big(\frac{v_S}{\sqrt{2} \Lambda} 
\Big)^{n^{e}_{ij}}\,\frac{v_H}{\sqrt{2}} \equiv Y^{(e)}_{ij} \,\frac{v_H}{\sqrt{2}} ~.
\end{eqnarray}
The SM Yukawa couplings $Y_{ij}^{(f)}$ 
are then hierarchical naturally for $f=(u,d,e)$, expressed in powers of a small expansion parameter $\epsilon$ defined as
\begin{equation}
    \epsilon \equiv \frac{v_{S}}{\sqrt{2} \Lambda}~.
\end{equation}
For a given set of fermionic $U(1)_{\rm FN}$ charges, a preferred value of $\epsilon$ can be identified that makes the hierarchies in fermion masses the CKM mixing angles natural, in the sense that the Yukawa couplings $y_{ij}^{(f)}$ of Eq. (\ref{eq:FN}) will be all of the order unity.  Since the largest of small numbers among fermion mass ratios and the CKM mixing angles is the Cabibbo angle, it is reasonable to take $\epsilon \simeq 0.225$, which is what we shall assume in our numerical studies. Note that the SM Yukawa couplings are now given by 
\begin{equation}
    Y^{(u)}_{ij} = y^{(u)}_{ij} \epsilon^{n^u_{ij}}, ~~~    Y^{(d)}_{ij} = y^{(d)}_{ij} \epsilon^{n^d_{ij}},~~~     Y^{(e)}_{ij} = y^{(e)}_{ij} \epsilon^{n^e_{ij}}~.
\end{equation}
For appropriate $U(1)_{\rm FN}$ charges of the fermions, which will be further discussed in Sec. \ref{sec:scen1}, and for $\epsilon = 0.225$, the hierarchical pattern of the SM Yukawa couplings is explained without assuming the Lagrangian parameters $y^{(f)}_{ij}$  to be hierarchical.  We now turn to generalize the FN mechanism to include a DM fermion.

\subsection{Extending $U(1)_{\rm FN}$ Flavor Symmetry to the Dark Sector}
\label{sec:dm}

We shall take the dark matter $\chi$ to be a Majorana fermion  which is a singlet of the SM, but charged under the $U(1)_{\rm FN}$ flavor symmetry.  The $U(1)_{\rm FN}$ charge of $\chi$ is denoted as $n$, which can take an integer or a half-integer value (see below). The interaction Lagrangian of $\chi$ is given  by 
\color{black}
\begin{eqnarray}
 \mathcal{L_{\rm DM}} &=& \frac{1}{2}\overline{\chi}\Big( i \gamma^\mu \partial_\mu \Big)\chi -  y_{\chi}\Big(\frac{S}{\Lambda}\Big)^{2n-1} S \,\overline{\chi^c} \chi + h.c.
 \label{LDM}
\end{eqnarray}
where $y_{\chi}$ is  an order one Yukawa coupling in the spirit of the Froggatt-Nielsen mechanism. 
This would lead to a dark matter mass given by
\begin{eqnarray}
 m_{\chi}&=& {\sqrt{2}} y_\chi v_S \epsilon^{2n-1}. 
 %\Rightarrow v_S = \frac{\sqrt{2} m_{\chi}}{y_\chi \epsilon^{2n-1}}
 \label{eq:vs}
\end{eqnarray}
 Note that the $U(1)_{\rm FN}$ charge is conserved in Eq. ({\ref{LDM}}) with $a_\chi = n$ and $a_S = -1$ for the charges of $\chi$ and $S$. 

There are several interesting features of the flavon-induced mass generation for the DM candidate $\chi$.  First, the coupling between the DM and the SM sectors arises primarily through the flavon portal.  Because of the suppressed nature of these couplings, the DM is naturally a FIMP candidate. Second, a bare mass for $\chi$ is not allowed by the $U(1)_{\rm FN}$ symmetry, and thus the mass of $\chi$ is protected, say compared to the Planck scale.  This model thus belongs to a class of chiral DM models \cite{deGouvea:2015pea,BhupalDev:2016gna,Berryman:2017twh,Abe:2019zhx} where a bare mass term is forbidden by symmetries.  Third, note that $n$, the charge of $\chi$, can take a half-integer value, in a normalization where the charge of $S$ is $-1$,  consistent with the mass generation for $\chi$.  If all the SM fermions have integer charges, $\chi$ cannot mix with the neutral fermions of the SM sector such as $\nu_L$ and $\nu_R$, even after $S$ acquires a VEV.  The stability of $\chi$ is therefore guaranteed by the $U(1)_{\rm FN}$ symmetry. As a benchmark value, we choose $n=15/2$, although other values will also be considered.  Such a charge for $\chi$ is analogous to the $U(1)_{\rm FN}$ charges assigned to the lighter generations of quarks and leptons in order to explain the fermion mass hierarchy (e.g., a charge of $+ 4$ for the first generation fermion of SM, see Sec. \ref{sec:scen1}), except for the fact that $n$ is a half-integer for $\chi$, while the charges of SM fermions are integers.  It is tempting to think of a whole new dark sector with many fermions, all neutral under the SM gauge symmetry, with $U(1)_{\rm FN}$ charge assignment similar to the SM sector, but shifted by half an integer unit.  In such a setup the lightest of the dark sector fermions will be naturally a dark matter candidate. Although the dark sector with multiple fermions is an intriguing possibility, for our analysis in this paper we shall stick with a single DM particle $\chi$.

The interaction Lagrangian involving the DM and the flavon field of Eq. (\ref{LDM}) can be expanded as:
\begin{eqnarray}
&y_\chi&  \Big(\frac{S}{\Lambda}\Big)^{2n-1} S\ \overline{\chi^c} \chi = \Big(\frac{h_S+v_S+i A_S}{\sqrt{2}\Lambda}\Big)^{2n-1} \Big(\frac{h_S+v_S+i A_S}{\sqrt{2}}\Big) \overline{\chi^c} \chi  \nonumber \\
 &=& \frac{y_\chi}{\sqrt{2}} \epsilon^{2 n-1}  \Big(v_S + 2 n ~(h_S+i A_S)~ +\frac{n\,(2n-1)}{v_S}(h^2_S+ 2i h_S A_S -A^2_S)\Big)\overline{\chi^c} \chi~~~[\rm{Before~EWSB}
 ] \nonumber \\
 &=& \frac{y_\chi}{\sqrt{2}} \epsilon^{2 n-1}  \Big(v_S + 2n ~(h_1 \sin{\theta}+h_2 \cos{\theta}+i A_S)~+...\Big)\overline{\chi^c} \chi ~~~~[\rm{After~EWSB}]
 \end{eqnarray}
 where in the second line we have dropped terms with powers larger than two in the fields.  These interactions will be used in the numerical study of the DM relic abundance.  We now turn to specific flavor models based on the FN mechanism.

\subsection{Flavon Sceranio I}
\label{sec:scen1}
In order to explain the observed hierarchical pattern of quark and lepton masses and mixings, we shall  adopt two specific flavor models based on the Froggatt-Nielsen mechanism. In the first model, the FN charges, with the normalization chosen such that the charge of $S$ is $-1$ and the SM Higgs doublet $H$ is neutral, are assigned as follows \cite{Babu:2003zz}:

\begin{equation}
\begin{pmatrix}
a_{Q_1} & a_{Q_2} & a_{Q_3} \\
a_{u_1} & a_{u_2} & a_{u_3} \\
a_{d_1} & a_{d_2} & a_{d_3} \\
a_{L_1} & a_{L_2} & a_{L_3}\\
a_{e_1} & a_{e_2} & a_{e_3} 
\end{pmatrix}\,
=
\begin{pmatrix}
4 & 2 & 0 \\
4 & 2 & 0 \\
4 & 3 & 3 \\
4 & 3 & 3 \\
4 & 2 & 0
\end{pmatrix}\,.
\label{eq:charge}
\end{equation}
Note that such a choice of charges is compatible with $SU(5)$ unification, since the charges of $(Q_i,\,u^c_i,\,e^c_i)$ fields which form $10_i$ multiplets of $SU(5)$ are all the same for each generation $i$, as are the charges of $(d^c_i,\,L_i)$ which form  $\bar{5}_i$ multiplets. 
Inserting this charges into Eq. (\ref{eq:FN}) one obtains the mass matrices for the up-quarks, down-quarks and charged leptons as:
\begin{align} 
	& M_u \sim \langle H \rangle \begin{pmatrix}
		\epsilon^8 & \epsilon^6 & \epsilon^4    \\
		\epsilon^6 & \epsilon^4 & \epsilon^2
		   \\
		\epsilon^4 & \epsilon^2 & 1 
	\end{pmatrix}   
,\qquad M_d \sim  \langle \widetilde H \rangle \epsilon^3 \begin{pmatrix}
		\epsilon^5 & \epsilon^4 & \epsilon^4   \\
		\epsilon^3 & \epsilon^2 & \epsilon^2
		   \\
		\epsilon & 1 & 1 
	\end{pmatrix}   
	& M_e \sim \langle \widetilde H \rangle \epsilon^3 \begin{pmatrix}
		\epsilon^5 & \epsilon^3 & \epsilon    \\
		\epsilon^4 & \epsilon^2 & 1
		   \\
		\epsilon^4 & \epsilon^2 & 1 
	\end{pmatrix}   ~.
 \label{eq:mat1}
 \end{align}
Here each entry in the matrices has an order one Yukawa coupling coefficient which is not explicitly shown.  These matrices, which are modifications of the lop-sided matrices proposed in the context of $SU(5)$ unified theory to explain simultaneously the small CKM mixing angles and the large neutrino mixing angles \cite{Babu:1995hr,Sato:1997hv,Irges:1998ax}, can provide a good understanding of several intricate features of the fermion mass and mixing pattern.  The masses obey the following hierarchy:
\begin{eqnarray}
m_u:m_c:m_t \sim \epsilon^8:\epsilon^4:1,~~~m_d:m_s:m_b &\sim& \epsilon^5:\epsilon^2:1,~~~m_e:m_\mu:m_\tau \sim \epsilon^5:\epsilon^2:1 \nonumber \\
m_\tau:m_b:m_t &\sim& \epsilon^3:\epsilon^3:1~.
\end{eqnarray}
For $\epsilon = 0.225$ and for the order one coefficients lying in the range $(0.3-1.2)$ all the observed fermion masses can be explained \cite{Babu:2004th}.  As for the CKM mixing angles, they obey the pattern
\begin{equation}
|V_{us}| \sim \epsilon^2,~~~|V_{cb}| \sim \epsilon^2,~~~|V_{ub}| \sim \epsilon^4~,
\end{equation}
along with $\delta_{CKM} \sim {\cal O}(1)$, 
which also provides a good fit to data (although the coefficient on $|V_{us}|$ should be slightly larger than unity). Neutrino masses and mixings can be readily incorporated into this framework, which would lead to large leptonic mixing angles and a weaker hierarchy in the neutrino masses.  Note that the form of $M_d$ and $M_e$ differ in a transposition in Eq. (\ref{eq:mat1}), which helps to realize large leptonic mixing angles arising from $M_e$.  If we adopt an effective set of $d=5$ Weinberg operators for neutrino mass generation or invoke the seesaw mechanism with $\nu^c$ fields which all have the same $U(1)_{\rm FN}$ charge, the form of the light neutrino mass matrix would be
\begin{align} 
 M_{\nu}^{light} \sim \frac{\langle H \rangle^2}{M_R} \epsilon^{6} \begin{pmatrix}
		\epsilon^2 & \epsilon & \epsilon    \\
		\epsilon & 1 & 1
		   \\
		\epsilon & 1 & 1  
	\end{pmatrix} ~.
\end{align}
This form is consistent with large mixing in the $\mu-\tau$ sector,  non-negligible mixing in the $e-\mu$ sector, along with smaller mixing in the $e-\tau$ sector.  A statistical analysis of this framework where the order one coefficients are allowed to take random values of order unity shows a good quality of the overall fit to fermion masses and mixings \cite{Babu:2016aro}. 

With the choice of $U(1)_{\rm FN}$ charges as shown in Eq. (\ref{eq:charge}), which lead to the mass matrices of Eq. (\ref{eq:mat1}) the most significant higher dimensional operators involving the flavon field are:
\begin{eqnarray}
{\cal L}_{\rm SM}^{\rm FN-Yuk} &\supset& 
\left(y_{23}^{(u)} Q_2\, u_3^c + y_{32}^{(u)}\,Q_3\, u_2^c\right)\,\left(\frac{S}{\Lambda}\right)^2H + \left(y_{32}^{(d)}\, Q_3 d_2^c + y_{33}^{(d)} \, Q_3 \,d_3^c\right) \left(\frac{S}{\Lambda}\right)^3\tilde{H} \nonumber \\
&+& \left(y_{23}^{(e)}\, L_2 e_3^c + y_{33}^{(e)}\, L_3 \,e_3^c\right)\left(\frac{S}{\Lambda}\right)^3\tilde{H} + h.c.
\label{eq:power}
\end{eqnarray}
 In particular, the scattering  $f +\overline{f} \rightarrow H + S$ and processes related to this via crossing symmetry play a crucial role in keeping the flavon in thermal equilibrium with the plasma in the early universe at temperatures below the $U(1)_{\rm FN}$ symmetry breaking scale, but above the EWSB scale. In order to determine the strengths of the relevant couplings, we partially diagonalize the down-quark and charged lepton interaction terms by defining
 \begin{equation}
 \hat{d}_3^c = 
\frac{y_{32}^{(d)} d_2^c + y_{33}^{(d)} d_3^c}{\sqrt{\left|y_{32}^{(d)}\right|^2 + \left|y_{33}^{(d)}\right|^2}},~~~ \hat{L}_3 = 
\frac{y_{23}^{(e)} L_2 + y_{33}^{(e)} L_3}{\sqrt{\left|y_{23}^{(e)}\right|^2 + \left|y_{33}^{(e)}\right|^2}}
 \end{equation}
 in which case $\hat{d}_3^c$ and $\hat{L}_3$ would be the approximate mass eigenstates, while the top-quark is already in the mass eigenstate in Eq. (\ref{eq:power}). The mass terms and interaction terms linear in the $h_S$ field resulting from Eq. (\ref{eq:power}) are then
 \begin{eqnarray}
    {\cal L}_{\rm SM}^{\rm FN-Yuk} &\supset& m_t\, u_3 u_3^c + m_t \epsilon^2 \left(a_{23}^{(u)} Q_2 u_3^c + a_{32}^{(u)} Q_3 u^c_2\right)\left\{ \frac{2 h_S}{v_S} + \frac{2 h_S\, h_H}{v_S\, v_H}\right\} \nonumber \\
    &+& m_b\, d_3\hat{d}_3^c + m_b \left(Q_3 \hat{d}^c_3\right) \left\{\frac{3h_S}{v_S} +\frac{3h_S\,h_H}{v_S\,v_H} \right\} \nonumber \\
    &+& m_\tau\, e_3\hat{e}_3^c + m_\tau \left(L_3 \hat{e}^c_3\right) \left\{\frac{3h_S}{v_S} +\frac{3h_S\,h_H}{v_S\,v_H} \right\} + h.c.
    \label{eq:int2}
 \end{eqnarray}
 Here we have identified the third family of fermion masses to be
 \begin{eqnarray}
     m_t = y_{33}^{(u)} \left(\frac{v_H}{\sqrt{2}}\right),~ m_b = \sqrt{\left|y_{32}^{(d)}\right|^2 + \left|y_{33}^{(d)}\right|^2} \epsilon^3 \left(\frac{v_H}{\sqrt{2}}\right), ~
     m_\tau = \sqrt{\left|y_{23}^{(e)}\right|^2 + \left|y_{33}^{(e)}\right|^2} \epsilon^3 \left(\frac{v_H}{\sqrt{2}}\right)~~~~~~
 \end{eqnarray}
 and defined $a_{ij}^{(u)} \equiv y_{ij}^{(u)}/y_{33}^{(u)}$, which should take values of order unity. The interaction terms of Eq. (\ref{eq:int2}) would be relevant for flavor violation mediated by the $h_S$ field, as well as for the process $f + \overline{f} \rightarrow h_S + h_H$ which would determine the freeze-in condition of the flavon field in the early universe.  (In the electroweak symmetric limit one should replace $h_H/\sqrt{2}$ by the doublet field $H$ in Eq. (\ref{eq:int2}).) For our numerical study we shall choose the running mass parameters evaluated at $\mu = 1$ TeV as $m_b(1~{\rm TeV)} = 2.43$ GeV, $m_\tau(1~{\rm TeV}) = 1.78$ GeV and $m_t(1~{\rm TeV}) = 151$ GeV \cite{Babu:2009fd,Xing:2007fb,Das:2000uk}. Furthermore, the couplings $a_{23}^{(u)}$ and $a_{32}^{(u)}$ of Eq. (\ref{eq:int2}), which are not entirely determined from the fermion masses or mixing angles, will be assumed to take values close to 1.0.  

\subsubsection{Flavon-induced flavor violation}

We shall be interested in a scenario where the components of the flavor field $S$, viz., $h_S$ ad $A_S$, have masses of order 1 TeV.  Here we show that such a choice is consistent with constraints from flavor-changing neutral current processes mediated  by these fields at the tree level.  To see this, we write the coupling matrices of the flavon fields in the fermion mass eigenbasis as follows:
\begin{equation}
    {\cal L}_{\rm FCNC}^{S} = a_{ij}^{(f)} f_i f^c_j (h_S+i A_S) + h.c.
\end{equation}
for $f = (u,d,e)$.  The flavon Yukawa coupling matrices in the fermin mass eigenbasis are:
\begin{eqnarray}
    a^{(u)} = \frac{1}{v_S}\left(\begin{matrix}  8 m_u & \epsilon^2 m_c & \epsilon^4 m_t \\
    \epsilon^2 m_c & 4 m_c & \epsilon^2 m_t \\  \epsilon^2 m_t &  \epsilon^2 m_t  & \epsilon^4 m_t
    \end{matrix}\right),
        a^{(d)} = \frac{1}{v_S}\left(\begin{matrix}  8 m_d & \epsilon^2 m_s & \epsilon^4 m_b \\ \epsilon m_s & 5 m_s & \epsilon^2 m_b \\  \epsilon m_b &   \epsilon^2 m_b  & 3 m_b
    \end{matrix}\right),
     a^{(e)} = \frac{1}{v_S}\left(\begin{matrix}  8 m_e & \epsilon^2 m_\mu & \epsilon^4 m_\tau \\ \epsilon m_\mu & 5 m_\mu & \epsilon^2 m_\tau \\  \epsilon m_\tau &   \epsilon^2 m_\tau  & 3 m_\tau
    \end{matrix}\right).~~~~~~
    \label{eq:flavcoup}
\end{eqnarray}
Here the coefficients shown in the diagonal entries are exact, except for the (3,3) entry of $a^{(u)}$ which has an order one factor that is not shown, and all the off-diagonal entries also have order one factors that are not shown.  These order one coefficients are not as large as the $8$ that show up in the diagonal entries, since mass matrix diagonalization also partially diagonalizes the flavon coupling matrix. For example, $a_{32}^{(d)}$ is not of order one as might be inferred naively from Eq. (\ref{eq:mat1}), but is suppressed by $\epsilon^2$.  For a discussion of flavon-induced FCNC in a similar FN model see Ref. \cite{Bauer_2016}. 

 The most stringent limits on flavon-mediated flavor change would arise from processes such as $K^0-\overline{K^0}$ mixing which receive new tree-level contributions.  The effective Hamiltonian for $\Delta F =2$ meson-anitmeson mixing can be readily written down after integrating the $h_S$ and $A_S$ fields:
\begin{equation}
    H_{\rm eff} = -\frac{1}{2 M_k^2}\left( \overline{q}_i\left[ Y_{ij}^k \frac{1-\gamma_5}{2}+ Y_{ji}^{k*} \frac{1+\gamma_5}{2} \right]q_j\right)^2.
\end{equation}
Here $k=(h_S,\,A_S)$ and $Y_{ij}^k = a_{ij}$ for $k=h_S$ and $Y_{ij}^k = i a_{ij}$ for $k = A_S$.  We can now estimate the constraints on the flavor-changing flavon couplings from  the mass splitting of $(K^0-\overline{K^0})$, $(B^0_d-\overline{B^0_d})$, $(B^0_s-\overline{B^0_s})$ and $(D^0-\overline{D^0})$ systems. We follow the procedure outlined in Ref. \cite{Babu:2018uik} and make use of the QCD renormalization factors worked out there. The transition matrix element $M_{12}^\phi$ for a meson system $\phi$ composed of $\overline{q}_i q_j$ can be written down as 
\begin{eqnarray}
M_{12}^\phi = \langle \phi |H_{\rm eff} |\overline{\phi} \rangle = -\frac{f_\phi^2 m_\phi}{2 M_k^2}\left[  -\frac{5}{24}\frac{m_\phi^2}{(m_{q_i}+m_{q_j})^2} \left(Y_{ij}^{k^2} + Y_{ji}^{k*^2}\right).\,B_2.\,\eta_2(\mu) \right.
\nonumber \\
\left. + Y_{ij}^k Y_{ji}^{k*} \left(\frac{1}{12} + \frac{1}{2}\frac{m_\phi^2}{(m_{q_i} + m_{q_j})^2}\right).\,B_4.\, \eta_4(\mu)   \right]
\end{eqnarray}
where vacuum saturation and factorization formalism has been adopted, with the hadronic matrix elements (obtained from lattice calculations) denoted by $B_i$ and the QCD renormalization factors to go from the TeV scale to the hadron mass scale $\mu$ denoted as $\eta_i(\mu)$.  The numerical values of the $B$ factors are summarized in Ref. \cite{Bauer:2009cf}; we choose $(B_2,\,B_4) = (0.66,\,1.03)$ for the $K^0$ system, $(0.82,\,1.16)$ for the $B_d^0$ and $B_s^0$ systems, and $(0.82,\,1.08)$ for the $D^0$ system. We also use the $\eta(\mu)$ factors computed in Ref. \cite{Babu:2018uik} as
$\{\eta_2(\mu),\,\eta_4(\mu)\} = (2.55,\,4.36)$ for the $K^0$ system, $(1.89,\,2.82)$ for the $B_{d,s}^0$ systems, and ($2.17,\,3.62)$ for the $D^0$ system.  

For the $K^0$ system, using $m_K = 498$ MeV, $f_K = 160$ MeV, $m_s = 100$ MeV, and $m_d = 5$ MeV, if we demand that the new flavon-induced contributions to $\Delta m_K$ saturates the experimental value of $3.48 \times 10^{-12}$ MeV, we obtain a constraint $|Y_{sd}| \leq 2.8 \times 10^{-4}$ for $M_{h_S} = 1$ TeV. Note that there could be some cancellations between the $h_S$-induced flavor violation and those induced by $A_S$, but a precise cancellation would be unnatural since that would require the masses of $h_S$ and $A_S$, which are unrelated, to be equal. In our estimate quoted above, we kept only the $h_S$ contribution.   Here we also assumed that $Y_{ds} = \epsilon Y_{sd}$, with $\epsilon = 0.23$ which follows from Eq. (\ref{eq:flavcoup}), and took all couplings to be real.  This constraint on $|Y_{sd}|$ should be compared with the expectation from Eq. (\ref{eq:flavcoup}), which has $|Y_{sd}| = |\epsilon m_s/v_S|$.  For $v_S = 1$ TeV, the prediction of the model is $|Y_{sd}| \simeq 2.3 \times 10^{-5}$, which is about an order of magnitude below the experimental limit from $\Delta m_K$.  

As for the CP-violating parameter $|\epsilon_K| \simeq |{\rm Im}(M_{12}^K)|/(\sqrt{2}|\Delta m_K|) = 2.23 \times 10^{-3}$, we find that the new tree-level flavon-induced contributions would overshoot this value unless the condition $|Y_{sd}| \leq 2.2 \times 10^{-5}$ is satisfied, assuming maximal value for the phase.  Here again we chose $Y_{ds} = \epsilon Y_{sd}$. This constraint does set a limit on $v_S \geq 1$ TeV, corresponding to $m_{h_S} = 1$ TeV. We shall take this constraint into account in our numerical studies.  Thus we conclude  that choosing the flavon field to have a mass of 1 TeV is consistent with $K^0-\overline{K^0}$ mixing, even for a relatively low value of its VEV,  $v_S \sim 1$ TeV. 

For the $B_d^0-\overline{B_d^0}$ system, we repeat the analysis using $m_{B_d} = 5.28$ GeV, $f_{B_d} = 240$ MeV, and $\Delta m_{B_d} = 3.12 \times 10^{-13}$ GeV and obtain, under the condition that $Y_{db} = \epsilon^3 Y_{bd}$ (cf. Eq. (\ref{eq:flavcoup})), with both couplings being real, the constraint  $|Y_{db}| \leq 1.5 \times 10^{-3}$ corresponding to $m_{h_S} = 1$ TeV.  Comparing with the prediction in the flavon model given in Eq. (\ref{eq:flavcoup}), viz., $|Y_{bd}| = |\epsilon m_b/v_S| \simeq 5.6 \times 10^{-4}$, we see that a value of $v_S = 1$ TeV is consistent with $B_d-\overline{B_d}$ mixing. Once this constraint is satisfied, the $B_s-\overline{B_s}$ mixing constraint is automatically satisfied in the model, since the flavon contributions to $B_s$ mixing is smaller by a factor of $\epsilon$ compared to those to the $B_d$ mixing, and since the measured mass splitting $\Delta m_{B_s} = 1.17 \times 10^{-11}$ GeV is about a factor of 37 larger than $\Delta m_{B_d}$.  

For the $D^0-\overline{D^0}$ system we use $m_D = 1.864$ GeV, $f_D = 200$ MeV, $m_u = 4$ MeV, $m_c = 1.27$ GeV, and saturate the flavon-induced contribution with the experimental value of $\Delta m_D = 6.25 \times 10^{-15}$ GeV.  Using $Y_{uc} = Y_{cu}$, and with $h_S$ mass of 1 TeV, we obtain the constraint $|Y_{uc}| \leq 1.7 \times 10^{-4}$, to be compared with the model prediction $|Y_{uc}| = |\epsilon^2 \,(m_c/v_S)| \simeq 6.7 \times 10^{-5}$ (for $v_S = 1$ TeV).  We thus see that all flavor violation constraints from the various meson systems are satisfied for $m_{h_S} = 1$ TeV and with $v_S = 1$ TeV.

The flavor violation induced by tree-level exchange of $h_S$ and $A_S$ fields in the leptonic sector are all well below the current experimental limits.  For example, the decay $\mu \rightarrow 3 e$ has a decay amplitude given by
\begin{equation}
A(\mu \rightarrow 3 e) = -\frac{a_{11}^{(e)} a_{21}^{(e)*}}{4 m_{h_S}^2} \left[ 
  (\overline{e}_L \gamma_\mu \mu_L) (\overline{e}_R \gamma^\mu e_R) + 2 (\overline{e}_R e_L) (\overline{e}_R \mu_L)\right] + h.c.
\end{equation}
This leads to the branching ratio given by \cite{Kuno:1999jp}
\begin{equation}
    {\rm Br}(\mu \rightarrow 3 e) = \frac{1}{64\, G_F^2 \,m_{h_S}^4} |a_{11}^{(e)} a_{22}^{(e)}|^2~.
\end{equation}
Demanding this to be $\leq 1.0 \times 10^{-12}$ leads to the constraint $|a_{11}^{(e)} a_{22}^{(e)}| \leq 9.3 \times 10^{-5}$ for $m_{h_S} = 1$ TeV. 
The expectation for this quantity in the FN framework is  $|a_{11}^{(e)} a_{22}^{(e)}| \sim (8\, \epsilon m_e\, m_\mu)/m_{h_S}^2 \sim 10^{-10}$ for $v_S = 1$ TeV (cf. Eq. (\ref{eq:flavcoup})), which is well within limits.  Similarly, other processes such as $\tau \rightarrow 3 \mu$  are also found to be highly suppressed.

\subsection{Flavon Scenario II}

In this section, we describe a second scenario that has an advantage over the model of Sec. \ref{sec:scen1}. Here the $U(1)_{\rm FN}$ flavor symmetry can be gauged.  Owing to the various triangle anomaly cancellation conditions that must be satisfied, it is not easy to gauge a flavor symmetry consistently (for attempts along this line see Ref. \cite{Chen:2008tc,Tavartkiladze:2011ex}). Here we suggest a relatively simple way of identifying $U(1)_{I_{3R}}$ as an anomaly-free flavor symmetry. While the resulting model can explain the main hierarchical features of the fermion masses, it does not fare as well as the previous model (cf. Sec. \ref{sec:scen1}) in addressing the intricate details of the fermion mass and mixing pattern.  Nevertheless, this model can also lead naturally to a Majorana fermion dark matter, which belongs to the FIMP category.  

With the inclusion of three right-handed neutrinos into the SM spectrum, it is well known that $U(1)_{I_{3R}}$ becomes an anomaly-free symmetry.  Under this $U(1)$, only the right-handed fermions carry charges, with $(u_R,\,d_R)$ carrying charge of $+1$ and $(\nu_R,\,e_R)$ carrying charge of $-1$.  Conventionally these charges are taken to be flavor universal.  However, since the triangle anomalies cancel generation by generation, it is possible to assign each generation a multiple of the charges quoted, making this symmetry a gauged flavor symmetry.  In order to employ the Froggatt-Nielsen mechanism in this context, we choose the $U(1)_{\rm FN}$ charges of fermion fields to be:
\begin{equation}
\begin{pmatrix}
a_{Q_1} & a_{Q_2} & a_{Q_3} \\
a_{u_1} & a_{u_2} & a_{u_3} \\
a_{d_1} & a_{d_2} & a_{d_3} \\
a_{L_1} & a_{L_2} & a_{L_3}\\
a_{e_1} & a_{e_2} & a_{e_3} \\
a_{\nu_1} & a_{\nu_2} & a_{\nu_3}
\end{pmatrix}\,
=
\begin{pmatrix}
0 & 0 & 0 \\
4 & 2 & 0 \\
-4 & -2 & 0 \\
0 & 0 & 0 \\
-4 & -2 & 0\\
4 & 2 & 0
\end{pmatrix}\,.
\label{eq:charge1}
\end{equation}
Here we have used a notation similar to the one in Eq. (\ref{eq:charge}), but with the $SU(2)_L$ singlet fields identified as right-handed, including the right-handed neutrino which is needed for anomaly cancellation.  Furthermore, we choose $a_H = 0$ and $a_S = -1$, as before, for the charges of the Higgs field $H$ and the flavon field $S$.  The Yukawa  Lagrangian for this set of charges is given by
\begin{eqnarray}
   \mathcal{L}_{\rm SM}^{Yuk} &\supset& \overline{Q}_{iL} u_{3R} \Tilde{H}+\overline{Q}_{iL} d_{3R} H+ \overline{\Psi}_{iL} e_{3R} H + \overline{\Psi}_{iL} \nu_{3R} \Tilde{H}\nonumber \\
   &+& \overline{Q}_{iL} u_{2R} \Tilde{H}\Big(\frac{S}{\Lambda}\Big)^2+\overline{Q}_{iL} d_{2R} H \Big(\frac{S^*}{\Lambda}\Big)^2+ \overline{\Psi}_{iL} e_{2R} H \Big(\frac{S^*}{\Lambda}\Big)^2+ \overline{\Psi}_{iL} \nu_{2R} \Tilde{H} \Big(\frac{S}{\Lambda}\Big)^2\nonumber\\
   & +& \overline{Q}_{iL} u_{1R} \Tilde{H}\Big(\frac{S}{\Lambda}\Big)^{4}+\overline{Q}_{iL} d_{1R} H \Big(\frac{S^*}{\Lambda}\Big)^{4}+ \overline{\Psi}_{iL} e_{1R} H \Big(\frac{S^*}{\Lambda}\Big)^{4}+ \overline{\Psi}_{iL} \nu_{1R} \Tilde{H} \Big(\frac{S}{\Lambda}\Big)^{4}.~~~~
\end{eqnarray}
The resulting mass matrices for the charged fermions take the form
\begin{equation}
M_{u,d,l,\nu}=\begin{pmatrix}
    \epsilon^4 & \epsilon^2 & 1 \\
     \epsilon^4 & \epsilon^2 & 1 \\
     \epsilon^4 & \epsilon^2 & 1  \\
    \end{pmatrix}m^0_{u,d,l,\nu}~.
    \label{eq:flav2}
\end{equation}
Here $M_\nu$ represents the Dirac neutrino mass matrix, and $\epsilon=\frac{v_S}{\sqrt{2}\Lambda}\simeq0.225$ is the small expansion parameter, as before. From here it follows that the mass eigenvalues are hierarchical, in the ratio $\epsilon^4: \epsilon^2:1$.  The CKM mixing angle hierarchy is not explained but has to be put in by hand in this modest scenario.  However, this scheme does have a reasonable set of charges for flavor, as given in Eq. (\ref{eq:charge1}), which are integers ranging in magnitude from 0 to 4.  (This is to be contrasted with the model of Ref. \cite{Chen:2008tc}, where the integer charges range in magnitude from 9 to 59.  With such flavor charges the model does explain many more features of the fermion masses and mixings.)  Anomaly cancellation via the Green-Schwarz mechanism \cite{Green:1984sg}, realized in the context of string theory, is more economical since the anomaly coefficients need not vanish in this case, but are simply related; however, this $U(1)$ symmetry would be broken at a scale close to the string scale, which may not be appropriate for low energy flavor models. For original FN models based on anomalous $U(1)$ flavor symmetry realized near the Planck scale, see  Ref. \cite{Ibanez:1994ig,Binetruy:1994ru}.

As for the Majorana masses of the $\nu_R$ fields, they are given by the Lagrangian
\begin{eqnarray}
   \mathcal{L}_{\rm Majorana } &=& M_{\nu_3}\left[ \nu_{3R} \,\nu_{3R}+  \nu_{3R} \nu_{2R} \Big(\frac{S}{\Lambda}\Big)^2  + \nu_{2R} \nu_{2R} \Big(\frac{S}{\Lambda}\Big)^4 +\right. \nonumber \\
 &~& \left.  + ~\nu_{3R} \nu_{1R}
   \Big(\frac{S}{\Lambda}\Big)^4 + \nu_{1R} \nu_{2R}
   \Big(\frac{S}{\Lambda}\Big)^6 + \nu_{1R} \nu_{1R}
   \Big(\frac{S}{\Lambda}\Big)^8 \right] + h.c.
\end{eqnarray}
where order one coefficients are not shown.  Defining a parameter $\tilde{\epsilon} \equiv v_S/M_{\nu_3}$, we can write down the resulting heavy Majorana neutrino mass matrix as
\begin{equation}
    M_R = \begin{pmatrix}
    \Tilde{\epsilon} \epsilon^3 & \Tilde{\epsilon} \epsilon^2 & \Tilde{\epsilon} \epsilon \\
    \Tilde{\epsilon} \epsilon^2 & \Tilde{\epsilon} \epsilon & \Tilde{\epsilon} \\
    \Tilde{\epsilon} \epsilon & \Tilde{\epsilon} & 1
    \end{pmatrix} 
    M_{\nu_3}~.
\end{equation}
The resulting light neutrino mass matrix, obtained from seesaw diagonalization with the Dirac neutrino mass matrix given in Eq. (\ref{eq:flav2}), can explain the observed features of the neutrino mass and mixing pattern.  

A fermionic dark matter candidate can be readily accommodated in this framework of gauged flavor $U(1)$.  Since the $U(1)$ symmetry is gauged, a pair of two-component fermions $\chi$ and $\bar{\chi}$ carrying $U(1)$ charges of $n$ and $-n$ are introduced, so that the theory remains anomaly free.  Here we shall choose $n$ to be an integer, so that $\bar{\chi}$ field becomes unstable, via its mixing with SM fields. We shall also assume a discrete $Z_2$ symmetry acting on $\chi$ field, but not the $\bar{\chi}$ field.  With this $Z_2$ symmetry, which remains unbroken, the stability of $\chi$ is guaranteed. A Dirac mass term connecting $\chi$ and $\bar{\chi}$ is not allowed by the $Z_2$, which implies that $\chi$ will acquire its mass via the flavon field, as in Eq. (\ref{LDM}).  For $n=6-8$ the dark matter phenomenology can remain similar to the case of the first scenario.  

As for the field $\overline{\chi}$, it could mix with the $\nu_R$ fields, with the dominant mass term arising from
\begin{equation}
    {\cal L}_{\bar{\chi}} = M_{\nu_3}\left(\bar{\chi}\,\nu_{3R}\right)\left(\frac{S^*}{\Lambda}  \right)^n~.
\end{equation}
For $n=6-8$, such a coupling would lead to a mass of $\bar{\chi}$ -- after a seesaw diagonalization -- $m_\chi \sim 10^{-8} M_{\nu_{3R}}$, which could be well above the TeV scale, depending on the mass scale $M_{\nu_{3R}}$. Hence $\bar{\chi}$ may be viewed as part of the SM, by virtue of this interaction and its mass. Thus $\overline{\chi}$ would not play a major role in the dark matter abundance calculation.

In a gauged $U(1)$ theory, a flavor gauge boson $X_\mu$ is present in place of the pseudoscalar field $A_S$ which appeared in the global $U(1)_{\rm FN}$ model of the previous section.  Its interactions could affect DM abundance calculations in general, but if the theory is only weakly gauged, these effects could be less important. This caveat should be kept in mind while applying the relic abundance calculation that we turn to in the next section. There we shall primarily focus on the global $U(1)_{\rm FN}$ scenario of Sec. \ref{sec:scen1}. A more natural scenario would be to have reheat temperature $T_R$ after inflation  less than the mass of the flavor gauge boson, so that $X_\mu$ is not thermalized post-inflation.  

While we have seen that the flavon-mediated flavor change would require the VEV of $S$ to be above 1 TeV for a flavon mass of 1 TeV, in the dark matter relic abundance calculation we will see that for the DM to be a FIMP, $v_S \gg m_{h_S}$ is preferred.  In this case, the FCNC constraints mediated by $h_S$ are easily satisfied, even for the flavon mass below 1 TeV. Such a scenario would require the quartic coupling $|\lambda_{HS}| \ll 1$ (cf. Eq. (\ref{eq:param})), which may appear unusual, but this choice is stable under radiative corrections, and thus technically natural.

%==============================================================================================================
\section{FIMP Dark Matter Phenomenology}
\label{sec:DMphen}

In this section, we shall analyze the dark matter phenomenology in the FN framework.  
\subsection{Model Parameters relevant for Dark Matter Phenomenology}
\noindent  The dark matter relic density  is mainly governed by the following independent parameters:
\begin{eqnarray}
\label{eq:i_par}
	\{v_{S},~~~y_\chi,~~~n({\rm half ~integer}),~~M_{S}, ~~\sin\theta \}.
\end{eqnarray}
For the convenience of the numerical analysis in this section, we have considered the CP even and CP odd scalars of the flavon  field to be degenerate($ m_{h_S} =m_{A_S}\equiv M_S$). We shall also choose $y_\chi$ to be of order one. Other important parameters of the model like the mass of DM, $m_\chi$ can be evaluated using the Eq. (\ref{eq:vs}) and  the cut-off scale $\Lambda$ parameterized as $\Lambda \simeq 3.14 v_S$ with $\epsilon=0.225$.
%==========================================================
\subsection{Theoretical and Experimental constraints }\label{sec:const}
%%%%%%%%%%%%%%%%%%%%%%%%%%%%%%%%%%%%%%%%%%%%%%%%%
Here we collect the various theoretical and experimental constraints that the model parameters should satisfy. These constraints will be imposed in the DM relic abundance calculation.

\noindent $\bullet$\textbf{ Stability of the Higgs potential:}
In order to maintain the stability of the vacuum, the quartic couplings of the scalar potential should obey the following bounded from below  conditions~\cite{Chakrabortty:2013mha}:
\begin{align}
 \lambda_{H} \geq 0,~~~~\lambda_{S} \geq 0 ~~~{\rm and}~~~\lambda_{H S} + 2\sqrt{\lambda_H \lambda_S} \geq 0 .
\end{align}

%%%%%%%%%%%%
\noindent $\bullet$\textbf{ Perturbative Unitarity:} 
%%%%%%%%%%%%
The perturbative unitarity constraints associated with the $S$-matrix corresponding to $2 \leftrightarrow  2$
scattering processes involving all particles in the initial and final states have been studied, leading to the conditions~\cite{Horejsi:2005da,Bhattacharya:2017fid}:
\begin{align}
   |\lambda_H| < 4 \pi,~~|\lambda_S| < 4 \pi,~~|\lambda_{HS}| < 8 \pi \nonumber \\
   |(3 \lambda_H + 2 \lambda_S) \pm \sqrt{2 \lambda_{HS}^2 + (3 \lambda_H - 2 \lambda_S)^2 }| < 8 \pi~~.
\end{align}
%%%%%%%%%%%%
\noindent $\bullet$\textbf{ Collider constraint on flavon mass:} 
%%%%%%%%%%%%
The bounds on a complex scalar singlet $S$, the flavon in our case, typically arise from its mixing with the SM Higgs boson arising from both theoretical and experimental constraints \cite{ATLAS:2015ciy}. The $W-$boson mass correction provides upper limit on the scalar-SM Higgs mixing angle, $|\sin\theta|$ as a function of heavy Higgs mass, $m_{h_2}$. In the mass range of $m_{h_2} \sim \{250-1000\}$ GeV, the upper bound on the mixing angle is $  0.22$~\cite{Robens:2016xkb}.
\color{black}
For the lower mass region, $m_{h_2} \lesssim 250$ GeV,  the direct Higgs search
and observed Higgs signal strength at collider provides an upper limit on the $\sin\theta $ as $|\sin\theta| \lesssim 0.25$ \cite{ATLAS:2016neq}. The limits from the electroweak precision observables are mild compared to the limits obtained from $W$-boson mass corrections \cite{Lopez-Val:2014jva}. In our analysis, we consider $\sin\theta \lesssim \mathcal{O}(10^{-2})$ for TeV order $m_{h_2}$. 

%=========
\noindent$\bullet$\textbf{{ Higgs invisible decay constraint:}}
In the present scenario, the DM fermion is taken to be lighter than the Higgs, i.e., $2 m_{\chi} < m_{h_1}$. In this case, the SM Higgs can decay into  two dark matter particles  which is strongly constrained from the measurement of Higgs invisible decay branching fraction. The ATLAS collaboration
\cite{ATLAS:2020cjb} has constrained this branching ratio, defined as
\begin{equation}
    Br(h_1 \rightarrow inv)=  \frac{ \sin^2\theta ~\Gamma(h_S \rightarrow \overline{\chi^c} \chi)}{\sin^2\theta ~\Gamma(h_S \rightarrow \overline{\chi^c} \chi)+~\cos^2\theta~\Gamma (h_H \rightarrow {\rm SM}~{\rm SM})}
\end{equation}
with $\Gamma(h_H \rightarrow {\rm SM}~{\rm SM}) \simeq 4.1 $ MeV
to be less than $13\%$.

%%%%%%%%%%%%%%%%%%%
\noindent$\bullet$\textbf{{ Dark matter relic abundance constraints:}}
The observed DM relic density is constrained by the latest PLANK data as $\Omega_{\rm DM} h^2=0.11933\pm 0.00091$ \cite{Planck:2018vyg}. In this work, we focus on sub-MeV FIMP DM candidates with naturally suppressed coupling to the visible sector. Due to its suppressed couplings, the DM-nucleon scattering cross-section is extremely small and lies well below the existing bounds from direct search experiments. However, the light (keV-scale) FIMP DM faces strong constraints from structure formation with the Lyman-$\alpha$ forest data placing a lower bound on DM mass to be $m_{\chi} \gtrsim 15$ keV \cite{Decant:2021mhj}. 

\begin{figure}[H]
\centering
\includegraphics[scale=0.6]{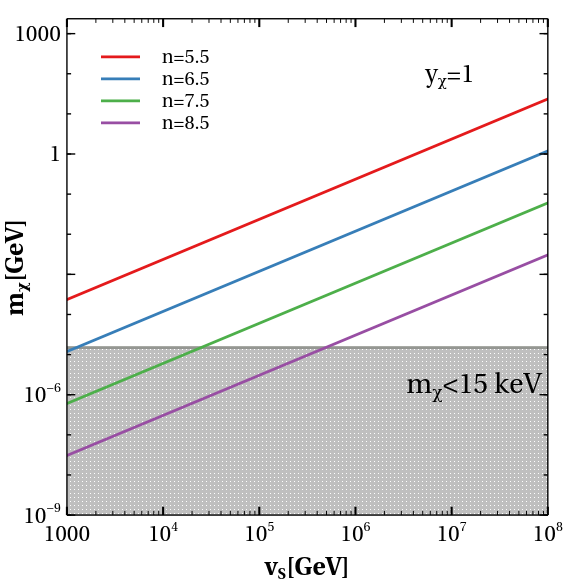}
 \caption{\it Calibration of $m_{\chi}$  as a function of $v_S$. Different colors correspond to different values of $n$. The lower gray shaded region is excluded from structure formation constraints. }
    \label{fig:calib}
\end{figure}

%%%%%%%%%%%%%%%%%%
\noindent$\bullet$\textbf{{ Constraint on flavon vev $v_S$:}}
The lower bound on the DM mass($\gtrsim 15$ keV \cite{Kamada:2019kpe}), imposes a constraint on the flavon vev $v_S$ which follows from Eq. (\ref{eq:vs}).
In Fig. \ref{fig:calib}, we present the calibration of $m_{\chi}$ as a function of $v_S$ for a fixed value $y_{\chi}=1.0$. The solid colored lines correspond to different values of $n$, the $U(1)_{\rm FN}$ charge of  DM field $\chi$.  The lower gray shaded region represents the DM mass constraint from structure formation \cite{Decant:2021mhj}. It is observed from the Fig. \ref{fig:calib} that for FN charges of DM $n=7.5$ and $n=8.5$, the allowed values of $v_S$ are larger than $\sim 2 \times 10^4$ GeV and $\sim 5 \times 10^5$ GeV  respectively. In our analysis we chose $v_S \gtrsim 10^6$ GeV for $n \lesssim 8.5$ to avoid  the excluded mass range of $m_\chi$.
%%%%%%%%%%%%%%%%%%%
\subsection{Dark Matter production through the freeze-in mechanism}
\label{sec:dmph}
%%%%%%%%%%%%%%%%%%%
We shall now discuss the production mechanism for the DM $\chi$. In the present scenario, the DM phenomenology is assumed to be conducted in a way similar to the FN mechanism with the coupling strength between the flavon and the DM controlling the DM production. We investigate the light DM case where its coupling to flavon is heavily suppressed due to the large FN charge of the DM. Therefore, the interactions between the DM particle and the flavon is so feeble that DM never attains thermal equilibrium in the early universe leading to the freeze-in scenario. The key feature for DM freeze-in~\cite{Hall:2009bx} production assumes that the DM starts with negligible initial abundance and gets populated with time through the decay of heavy particles present in the thermal bath.

%%%%%%%%%%%%%%%%%%%%%%%%%%%%%%%%%%%%%
After the $U(1)_{\rm FN}$ symmetry is broken, the DM $\chi$ is dominantly produced from the decays of flavon $S$ assumed to be in thermal equilibrium with the thermal bath particles. In this scenario, there are two types of interactions available to keep the flavon in the thermal bath: $(i)$ the dimension six Yukawa interaction $ y^{(u)}_{ij}\overline{Q}_{i} u^{c}_{j} \Tilde{H}\Big(\frac{S}{\Lambda}\Big)^2$ which leads to different kinds of  $2\leftrightarrow 3$ ($SSH \leftrightarrow q \overline{q}$ , $SS \leftrightarrow q \overline{q}H$) and $2 \leftrightarrow 2$ ($S H \leftrightarrow q \Bar{q}$ , $S q \leftrightarrow q H$) number changing scattering processes for S and $(ii)$ the Higgs portal interactions between the flavon and the Higgs, $(S^\dagger S)(H^\dagger H)$ which leads to the scattering process $SS \leftrightarrow H H$. Whether the flavon $S$ could be in thermal equilibrium or not depends on the interaction rates of the aforementioned scattering processes.  
The total interaction rate of $S$ at temperature $T$ ($T_{EW}<T<v_S$) is given by:
\begin{eqnarray}
{\bf \Gamma}_{S}&=&
{ \Gamma}_{2 \leftrightarrow 3}+ { \Gamma}_{2 \leftrightarrow 2} \approx { \Gamma}_{2 \leftrightarrow 2}  \nonumber \\
&=&{ \Gamma}_{S q \leftrightarrow H q}+{ \Gamma}_{S H \leftrightarrow \Bar{q}\, q}+{ \Gamma}_{S S \leftrightarrow H H} \nonumber \\
&=&  
n_q^{eq} {\langle \sigma v \rangle}_{S \, q\leftrightarrow H q }+n_H^{eq} {\langle \sigma v \rangle}_{SH\leftrightarrow q \Bar{q}}  +n_S^{eq} {\langle \sigma v \rangle}_{SS\leftrightarrow HH} ~.
\label{eqn:Sthermal}
\end{eqnarray}
\noindent The analytical expressions of the interaction rates for all scattering processes are presented in the appendix \ref{apx:IntRate}. The total interaction rate of $S$ ($\Gamma_S$) explicitly depends on the choice of
the flavon mass $M_{S}$, flavon VEV, $v_S$, and the Higgs mixing angle $\sin\theta$.    

Now, in order to check if any particle is in thermal equilibrium with other particles in the thermal bath, one needs to compare the interaction rate with the Hubble expansion rate,~$\mathcal{H}$. The conditions are as follows:
\begin{equation}
 \frac{{\bf \Gamma}}{\mathcal{H}}~~
    \begin{cases}
       > 1\,  &\text{: in thermal equilibrium}\,,\\[8pt]
    <1 &\text{: out of thermal equilibrium }\,.
    \end{cases}
\end{equation}
The Hubble expansion rate, $\mathcal{H}$, for standard radiation dominated universe is given by 
\begin{eqnarray}
\mathcal{H}&=& \sqrt{ \frac{ \pi^2 g_{\star\rho}}{90}}\,\frac{T^2}{M_{\rm Pl}}  ~~,\nonumber
\end{eqnarray}
where $g_{\star\rho}$ is the total relativistic degrees of freedom at temperature $T$ contributing to energy density and $M_{\rm Pl}= 2.4 \times 10^{18}$ GeV.

\begin{figure}
\centering
\includegraphics[scale=0.6]{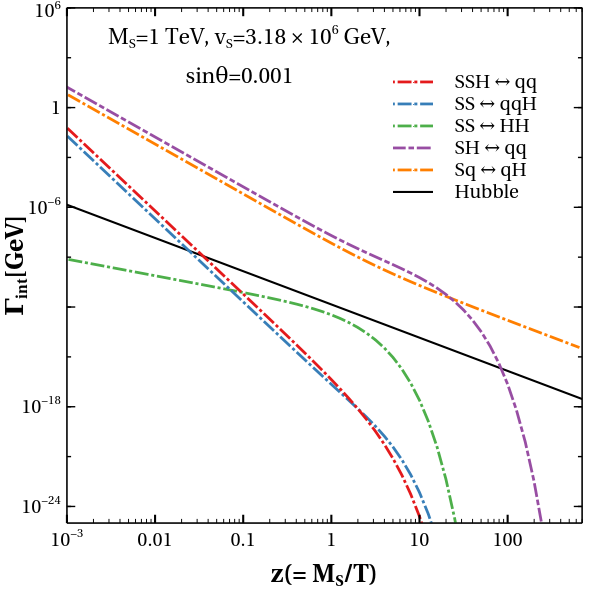}
 \caption{\it Plot shows the interaction rate of S as a function of $z=\frac{M_S}{T}$. The colored dashed lines correspond to the contribution from various processes aiding in the thermalization of $S$. The values of the chosen parameters are specified in the inset.}
    \label{fig:contributions}
\end{figure}
\begin{figure}
$$
\includegraphics[scale=0.5]{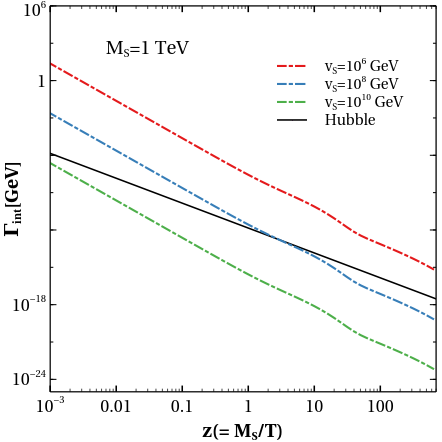}~~
\includegraphics[scale=0.51]{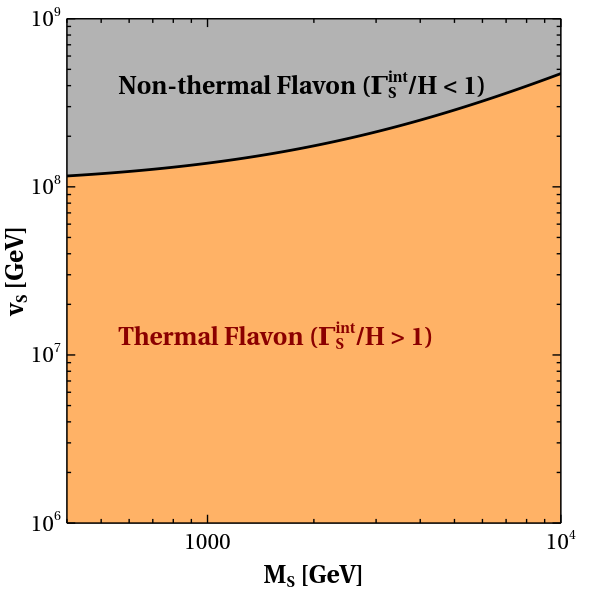}
$$
 \caption{\it [Left] The total interaction rate of $S$ and Hubble expansion rate are plotted as a function of  $\frac{M_S}{T}$ considering  the two dominant  $2 \leftrightarrow 2 $ number changing processes of $S$ ($SH \leftrightarrow q \Bar{q}$ and $Sq \leftrightarrow q H$ ) for different values of $v_S$ with $M_{S}=1$ TeV. [Right] 
 Thermalisation of flavon in $v_S$ vs $M_S$ plane. The orange area corresponds to a region where flavon is thermal i.e $\Gamma_S$ > $\mathcal{H}$ at $T=M_S$. Similarly the gray shaded region shows where $\Gamma_S$< $\mathcal{H}$ where the flavon is nonthermal.}
    \label{fig:Sthermal}
\end{figure}
 Let us now turn to the thermalization of the parent particle $S$ responsible for the DM production. In  Fig. \ref{fig:contributions} we present the interaction rate of $S$  with the thermal bath particles considering $3\leftrightarrow 2$ , $2\leftrightarrow 3$ and $2\leftrightarrow 2$ number changing processes as a function of $z=\frac{M_S}{T}$. The colored dashed lines correspond to the individual contributions coming from different processes that influences the interaction rate. Here the Hubble expansion rate is shown by the black solid line. The red and blue lines represent the contribution of $3 \leftrightarrow 2$ and $2 \leftrightarrow 3$ processes. For $T\gg M_{S}$, $SSH\leftrightarrow qq$ and $S S q\leftrightarrow q H$  contribute to the total interaction rate of $S$ which  decreases with the decrease of temperature, due to the large suppression of $({n_S^{eq}})^2$. But, $2 \leftrightarrow 2$ scatterings are always dominant over $2 \leftrightarrow 3$ processes. For the chosen temperature range shown, as $2 \leftrightarrow 3$ processes are subdominant compared to the $2 \leftrightarrow 2$ processes, we can neglect their contributions with respect to the $2 \leftrightarrow 2$ processes . The $2 \leftrightarrow 2$ processes  contribute dominantly to the thermalization of the flavon $S$. 
Among these $2 \leftrightarrow 2$ processes, the ones which arise from the dimension six operators i.e. $S q \leftrightarrow q H$ and $S H \leftrightarrow q \bar{q}$, influence the thermalization of the flavon most significantly. 
On the other hand, the value of $\lambda_{HS}$ (calculated from Eq. (\ref{eq:cplngs})), for the chosen values of $M_S=1$ TeV, $v_S=3.18\times 10^{6}$ GeV and $\sin\theta=0.001$ turns out to be $\lambda_{HS} \simeq 10^{-6}$, indicating that the Higgs portal coupling has a negligible contribution in thermalization of the flavon $S$. 
Here we see that the process $S q \leftrightarrow q H$ is dominant in lower temperatures compared to $S H \leftrightarrow q \bar{q}$ as light quarks have large enough number density to keep the processes going.

In Fig. \ref{fig:Sthermal} $[Left]$, we present the parameter dependencies of the interaction rates of $2 \leftrightarrow 2$ scattering processes in the thermalization of the flavon $S$. Here the total interaction rate is plotted as a function of $z= \frac{M_S}{T}$ for $M_S=1$ TeV. The different colors correspond to different values of $v_S$. The plot shows that the interaction rate decreases with an increasing value of $v_S$. This is because the interaction rates of $S$ depend on the effective coupling as $\sim \big(\frac{\epsilon}{3 v_S}\big)^2$. As we can see from the plots, it is possible to achieve better thermalization of $S$ for smaller values of $v_S$ (greater than $10^6$ GeV to be consistent with Lyman-alpha constraint for DM with $y_\chi=1$ and $n\lesssim 8.5$) where the interaction rate for flavon dominates over the Hubble expansion rate. Now, if the interaction rate of the flavon is smaller than the Hubble expansion rate at that temperature, then flavon will go out of equilibrium. But as here we are only interested in a scenario where flavon is thermal, we will find out a region where the interaction rate is greater than Hubble expansion rate. We will perform our analysis in that region. In Fig. \ref{fig:Sthermal}, in right panel, we show the thermalisation of $S$ in $v_S$ vs $M_S$ plane. $\Gamma_{S}$ is given by Eq. (\ref{eqn:Sthermal}). Now as the FN scale $\Lambda$ increases with $v_S$, the coupling of flavon and SM particles decreases. We observe that at $T= M_{S}$, flavon becomes non-thermal when the value of $v_S$ is roughly larger than $10^8$ GeV for $M_S$ in the range {$500 \rm~ GeV-10 \rm ~TeV$}.

\begin{figure}
\centering
\includegraphics[scale=0.6]{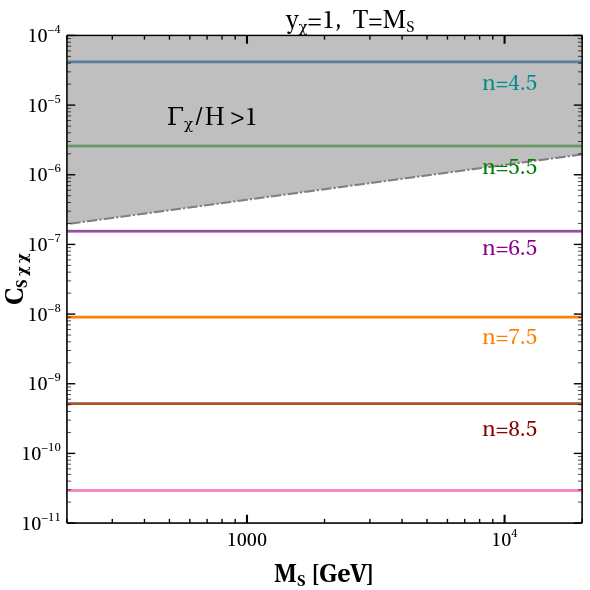}
 \caption{\it The plots are shown in $M_S-C_{S\chi\chi}$ plane where the effective coupling between flavon and DM is defined as $C_{S\chi \chi}=\frac{y_\chi 2n \epsilon^{2n-1}}{\sqrt2}$.  The solid horizontal lines correspond to possible flavon-DM coupling identified by $U(1)_{FN}$ charge of DM, $n$ where $\chi$ becomes a stable DM candidate. The  shaded region (above the dotted black line) is where the DM production from flavon falls in thermal equilibrium whereas the region below the gray dotted line fulfilled non-thermal DM production criteria from thermal flavon,  $({\Gamma}_{\chi} < \mathcal{H} )$.   
}
    \label{fig:NTDM}
\end{figure}

On the other hand, the DM $\chi$ can be produced non-thermally from the decay of the flavon which depends on the suppressed coupling, $\epsilon^{2n-1}$. The DM can gain thermal equilibrium through the tree-level interaction with our thermal flavon. The interaction rate is given by
  \begin{eqnarray}
  {\bf \Gamma}_{\rm \chi} 
  %&=& { \Gamma}_{\rm Dec}+{ \Gamma}_{\rm Scat}  \nonumber \\
  &=& \langle \Gamma_{S\to\chi\chi} \rangle+\langle \Gamma_{\rm SS \to \chi \chi} \rangle \nonumber \\
  &=&\Gamma_{S \to \chi\chi} \frac{K_1(M_{S}/T)}{K_2(M_{S}/T)}+ n^{eq}_S \langle \sigma v \rangle_{SS\to \chi \chi}
  \end{eqnarray}
  where $K_{i}$ denote the modified Bessel functions of the $i$th kind and
 $\Gamma_{S \to \chi\chi}$ is the decay width of the flavon to DM. $n^{eq}_S$ is the equilibrium number density of flavon and $\langle \sigma v \rangle_{SS\to \chi \chi}$ is the thermally averaged cross section of $SS \to \chi \chi$ process (analytical expression is given in Appendix \ref{apx:IntRate}). Note that as the CP even and CP odd states are chosen to be degenerate here, we have considered the number of degrees of freedom of flavon to be two.   
The analytical expressions of $\Gamma_{ {S} \to \chi\chi}$ has presented in Appendix \ref{apx:dw}. 
In Fig. \ref{fig:NTDM}, we show the interaction rate of $\chi$ in $M_{S}- C_{S \chi \chi}$ plane for the decay process $S\to \chi \chi$ and scattering process $SS \to \chi \chi$ assuming $S$ is in thermal equilibrium with the SM particles and $T=M_S$. The interaction rate depends on the parameters $n$ and $M_S$ and different values of $n$ have been represented by different colored lines. As the decay amplitude is proportional to $\epsilon^{2n-1}$, as $n$ increases, the interaction rate decreases accordingly. The solid line  corresponds to the values of $C_{S \chi \chi}$ for different values of $n$. The gray dashed line corresponds to the condition where $\Gamma^{int}_\chi/\mathcal{H}=1$. When the interaction rate is below the gray line, it indicates that $({\bf \Gamma} < \mathcal{H} )$ i.e. the DM particle produced from flavon decay goes out of equilibrium which represents our area of interest in this work. The gray shaded region is where $\chi$ follows  thermal equilibrium. We find that the DM $\chi$ remains non-thermal at high temperature  for higher values of $n$, i.e for $n \ge 5.5$ for $T=M_{S}$ and $y_{\chi}=1$.

%%%%%%%%%%%%%%%%%%%%%%%%%%%%%%%%%%%%%%%%%%%%%
\subsubsection{Boltzmann equations for the flavon and Dark Matter fields}
%%%%%%%%%%%%%%%%%%%%%%%%%%%%%%%%%%%%%%%%%%%%%
As discussed previously, due to the feeble coupling  which 
arises naturally in the FN mechanism, the dark matter never attains thermal equilibrium in this framework. However, it can be produced from the decay of the heavy flavon $S$ present in the thermal bath. In order to find the DM number density in this freeze-in scenario, one needs to solve two coupled Boltzmann equations for $S$ (degenerate $h_S$ and $A_S$) and $\chi$.  The Boltzmann equations are written in terms of the co-moving number densities $Y_\chi$ and $Y_{S}$  and are given by 
\begin{eqnarray}
 \frac{dY_\chi}{dz} &=& \frac{\langle \Gamma(S \to \chi \chi) \rangle}{\mathcal{H} ~ z}  Y_{\rm S}\big(z \big) +\frac{4\pi^2}{45} \frac{M_{Pl} M_{S}}{1.66}\frac{\sqrt{g_{\star(z)}}}{z^2} \langle \sigma v_{{\rm S \, S }\to \chi \,\chi} \rangle\,{Y^2_{\rm {S}}}(z) \nonumber \\
 \frac{dY_{S}}{dz} &=& -\frac{\langle \Gamma(S \rightarrow \chi \chi) \rangle}{\mathcal{H} ~ z}  Y_{\rm S}\big(z \big) -\frac{\langle \Gamma(S \rightarrow H \Bar{q} q) \rangle}{\mathcal{H} ~ z}  (Y_{\rm {S}}\big(z \big)-Y^{eq}_{\rm S}\big(z \big)) \nonumber \\
 &&-\frac{4\pi^2}{45} \frac{M_{Pl} M_{S}}{1.66}\frac{\sqrt{g_{\star(z)}}}{z^2} \langle \sigma v_{{\rm S \, S }\to \chi \,\chi} \rangle\,{Y^2_{\rm {S}}}(z)\nonumber \\
 &&-\frac{4\pi^2}{45} \frac{M_{Pl} M_{S}}{1.66}\frac{\sqrt{g_{\star(z)}}}{z^2} \langle \sigma v_{{\rm S \, q }\to q\,H} \rangle\,Y^{eq}_{\rm {q}}(z)({Y_{\rm {S}}}(z)-{Y^{eq}_{\rm {S}}}(z)) \nonumber \\
 && -\frac{4\pi^2}{45} \frac{M_{Pl} M_{S}}{1.66}\frac{\sqrt{g_{\star(z)}}}{z^2} \Big(\langle \sigma v_{{\rm S S}\to H H} \rangle\,({Y^2_{\rm {S}}}(z)-{Y^{eq}_{\rm {S}}}^2(z))\nonumber \\
 &&-\frac{4\pi^2}{45} \frac{M_{Pl} M_{S}}{1.66}\frac{\sqrt{g_{\star(z)}}}{z^2}\langle \sigma v_{{\rm S H}\to q \Bar{q}} \rangle\,Y^{eq}_{\rm {H}}(z)({Y_{\rm {S}}}(z) -{Y^{eq}_{\rm {S}}}(z)) \Big)~. 
\label{Boltzman}
\end{eqnarray}
 The dominant contribution to the DM density comes from the decay of $S$  since the scattering diagrams include propagators and additional couplings that are suppressed compared to the decay processes.
\iffalse 
 The thermally averaged decay-width for a given decay process, $A \to B+C$ can be defined as ~\cite{Babu:2014pxa}
\begin{equation}
  \langle \Gamma_{A \to B+ C} \rangle  = \big(\Gamma_{A \to B+ C}\big) \,\frac{K_1(m_A/T)}{K_2(m_A/T)}
\end{equation}
where $m_A$ is the mass of decaying particle and $K_1$ and $K_2$ are modified Bessel's functions. 
\fi

\begin{figure}
    \centering
\includegraphics[scale=0.6]{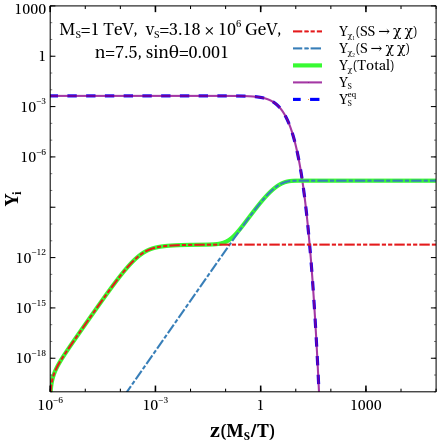} 
    \caption{\it Evolution of the DM comoving number density $Y_\chi$ (green thick line) and the flavon comoving number density $Y_S$ (magenta line) as a function of $z$($=\frac{M_S}{T}$). Here $M_S$ is the mass of the particle present in the SM thermal bath which is equal to $m_{h_S}=m_{A_S}$. The blue dashed line denotes the equilibrium yield $Y_{eq}$ for ${M_S}(=1 TeV)$. The values of other parameters chosen  for this plot is $v_S=3.18 \times 10^6  GeV$, $M_{S}=m_{h_2}=m_{A_S}$=1 TeV, $n=8.5$.}
    \label{fig:BEQ}
\end{figure}
We can now discuss the production of the DM through different decay channels and the role of different parameters affecting this scenario. For this one needs to solve the coupled BEQs for different choices of parameters: $\{v_S, ~n,~M_{S},~\sin\theta ,~y_{\chi}=1\}$.  In Fig. \ref{fig:BEQ}, we show the evolution of the DM number density with the dimensionless parameter $z=\frac{M_{S}}{T}$ with $M_{S}=1$ TeV. The dash-dotted blue (red) line corresponds to the co-moving number density of the DM when only the decay term (scattering term) is present in the number density evolution equation of $\chi$. As the decay term has a dimensionless coupling,  it naturally acts as a source of IR freeze-in like case whereas the scattering term with a dimensionful coupling can behave as a generator of UV freeze-in. The total co-moving number density of $\chi$ is represented by the green thick line. The navy blue dashed line corresponds to the equilibrium density of $S (h_S/A_S)$. The magenta solid line shows the evolution of $S$ co-moving number density obtained from solving the coupled Boltzmann equations. Initially, in the early universe, the density of the DM particle had negligible abundance. Then it started to get populated with decreasing temperature due to
the decay of parent particles as well as scattering until the abundance of parent particle ($S$) decreased due to the Boltzmann suppression for $T< M_S$. The DM has only non-thermal interaction with the flavon $S$ while the flavon is in the thermal bath via the dimension 6 operator and the Higgs portal interaction. So in the early universe, the DM is produced only through the tree-level interactions: decay process $S \to \chi \chi$ and scattering process $S S \to \chi \chi$. In order to see the evolution of the DM and the parent particles, we numerically solve the coupled Boltzmann equations given in Eq. (\ref{Boltzman}). We have also considered the effects of the EWSB in the scalar sector where $h_S$ mixes with the SM CP even Higgs giving rise to two physical states $h_1$ and $h_2$. For small $\sin\theta$, $h_2$ is nearly the same as $h_S$ with $m_{h_2} \simeq m_{h_S}$. Though many channels can open up after the EWSB through mixing, their contribution is negligible in the evolution of the DM in the small $\sin\theta$ limit. We only consider the dominant  processes shown in the Boltzmann equations of Eq. (\ref{Boltzman}).

From  Fig. \ref{fig:BEQ}, we can see that by solving the coupled Boltzmann equations, we get a solution for the co-moving number density of flavon $S$ which follows equilibrium number density. So instead of solving two coupled Boltzmann equations, we can simply solve a single  equation for $Y_{\chi}$ as
\begin{eqnarray}
   \frac{dY_\chi}{dz} &=& \frac{\langle \Gamma(S \rightarrow \chi \chi) \rangle}{\mathcal{H} ~ z}  Y^{eq}_{\rm S}\big(z \big) +\frac{4\pi^2}{45} \frac{M_{Pl} M_{S}}{1.66}\frac{\sqrt{g_{\star(z)}}}{z^2} \langle \sigma v_{{\rm S \, S }\leftrightarrow \chi \,\chi} \rangle\,{(Y^{eq}_{\rm {S}}})^2(z)
   \label{eqn:solve}
\end{eqnarray}
where $z=\frac{M_{S}}{T}$ and $Y_\chi$ is defined to be the ratio of the number density and the entropy density of the visible sector given by $Y_\chi=n_\chi/s$ for the DM. The equilibrium yield is given by
\begin{equation*}
Y_S^{eq}=\frac{45}{4\,\pi^4}\,\frac{g_{_S}}{g_{\star s}}\,(M_S/T)^2\,K_2(M_S/T)\,,    
\end{equation*}
 where $g_{_S}=2$ is the intrinsic number of degrees of freedom for the flavon S  and $g_{\star s}$ is the total number of relativistic degrees of freedom present in the thermal bath. Both the scattering and the decay contributions are present in the BEQ.

Note that although both the scattering and decay process contribute to the DM density, the scattering contributions are negligible in comparison to the decays due to a $v_S$ suppression in the coupling of the scattering processes. So the final relic density will always be controlled by the decay process. In the next section, we will calculate the relic density of our dark matter by following Eqn. \ref{eqn:solve} with the decay term only. 
% On the other hand, the Boltzmann equation for $S$ incorporates all possible $2 \leftrightarrow 2$ scattering processes($ S S \leftrightarrow HH,S S \leftrightarrow \chi \chi, S\,q \leftrightarrow q H, S \,H \leftrightarrow q \Bar{q}$) and decay processes involving $S \to \chi \chi$ and $S \to H q \Bar{q}$.
%%%%%%%%%%%%%%%%%%%%%%%
\subsection{DM parameter space from relic density calculation}
%======================================
The DM relic density is obtained via the non-thermal freeze-in mechanism obtained through the decay of the flavon. Here we discuss the role of various model parameters on relic abundance and present the allowed dark sector parameter space satisfying the DM observed density, $\Omega_{\rm DM} h^2=0.1200\pm0.001$, by PLANCK collaboration \cite{Planck:2018vyg}. 
In order to obtain the DM yields, $Y_\chi$, and hence DM abundance at the present time, we solve the sets of BEQs discussed earlier numerically. 
In this case, the decay $S \to \chi \chi$ is found to be the dominant production mode for the DM with the flavon $S$ being in a thermal bath throughout the DM production. Therefore the parameters involved in the decay process i.e the flavon mass, $M_S$, and the suppressed effective Yukawa coupling generated by the FN mechanism, $\epsilon^{2n-1}$ play an important role in setting the right relic abundance. The DM yield can be analytically expressed as \cite{Hall:2009bx} 
\begin{eqnarray}
    Y_\chi^{\rm today} &\approx& \frac{135 g_{_S}}{8\pi^3 (1.66) g_{*s} \sqrt{g_*}} \frac{M_{\rm Pl}}{M_S^2} \Big[ \frac{M_{S}}{16\pi}\Big(1-\frac{4 m^2_\chi}{M^2_{S}}\Big)^\frac{3}{2} ~\Big( \frac{y_\chi 2n \epsilon^{2n-1}}{\sqrt2}\Big)^2 \Big] \\
    && \propto \frac{\epsilon^{4n-2}~n^2}{M_S} \nonumber 
\end{eqnarray}
where $g_{*s} \simeq g_*$, $g_{_S}=2$  and $\epsilon \simeq 0.225$. Thus the abundance of the DM depends on the two independent parameters $M_S$ and $n$ for $m_\chi \ll M_S$.
\begin{figure}
\centering
   \includegraphics[width=7cm]{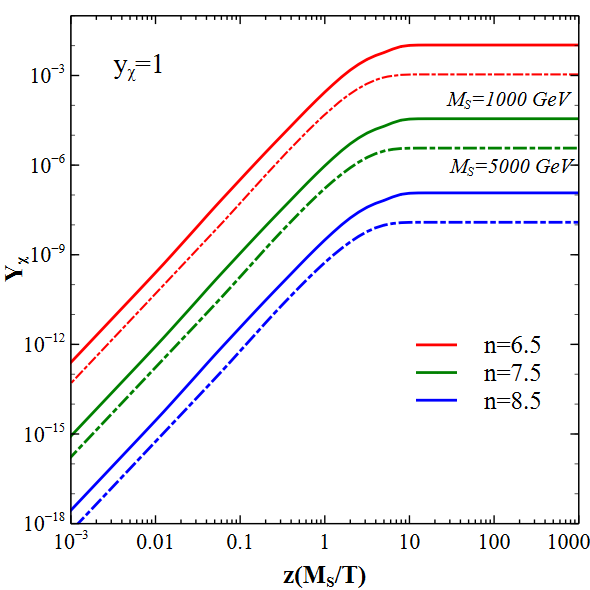}~~
   \includegraphics[width=7cm]{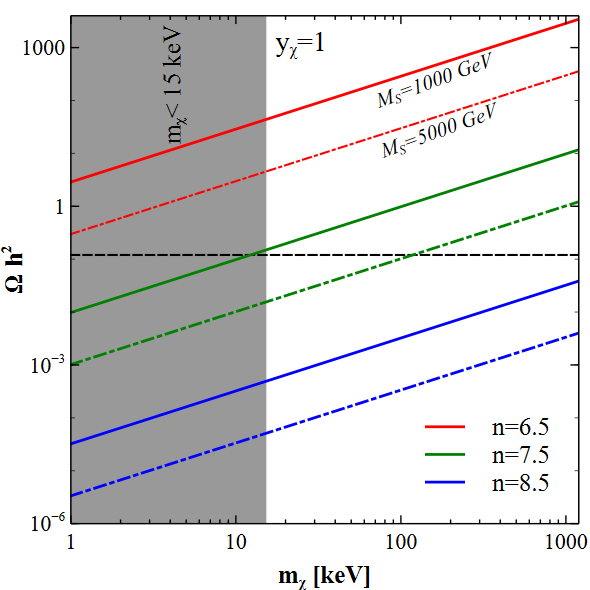}
\caption{\it [Left] Evolution of DM yields $Y_\chi$ as a function of $z=M_S/T$ for three different values of $n$: $n=6.5$ (red), $7.5$ (green), $8.5$ (blue). [Right] Relic density of DM as a function of $m_{\chi}$ with different choices of $n$: $n=$ 6.5 (red), 7.5 (green), and 8.5 (blue). Here DM mass $m_\chi$ is defined as $m_\chi={\sqrt{2}y_\chi v_S \epsilon^{2n-1}}$. Correct observed relic density bound measured by PLANCK is presented by the black dashed line. The gray-shaded region is excluded from the structure formation constraint (Lyman-$\alpha$ forest). Here the solid and dashed lines correspond to two different flavon masses $M_{S}=1$TeV, and $5$ TeV respectively. }
    \label{fig:6}
\end{figure}

In the left panel of Fig. \ref{fig:6} we present the evolution of the DM yields $Y_\chi$ with $z=M_S/T$ for different values of $n$ for a fixed flavon mass $M_S$. For illustration, we consider two different flavon masses $M_S= 1$ TeV and $5$ TeV which are depicted by solid and dot-dashed lines respectively. For example, for a given flavon mass, $M_S=1$ TeV (solid line), we choose three different values of $n$ i.e. $n=6.5$ (red), $7.5$ (green), and $8.5$ (blue). 
For a fixed $M_S$, with the increase of $n$, the effective Yukawa coupling between the DM and flavon $\epsilon^{2n-1}$ becomes more and more suppressed. As a result the co-moving number density for the DM, $Y_\chi$, decreases as $Y_\chi \propto \epsilon^{4n-2}$, which is  depicted in the figure. Similarly for a fixed value of $n$, with an increase of $M_S$, the decay width $\Gamma_{S\to\chi\chi}$ will decrease, leading to a decrease in $Y_\chi$ as $Y_\chi \propto 1/M_S$. This behavior is portrayed in the figure by the same color solid, and dot-dashed lines. It is important to note here that for $M_S \gg m_\chi$, the DM yield, $Y_\chi$, is almost independent of DM mass $m_\chi$.  In the right panel of Fig. \ref{fig:6} we show the variation of the DM abundance as a function of DM mass $m_\chi(\equiv m_{\rm DM})$. The density of DM has the following form
\begin{eqnarray}
    \Omega_\chi h^2 =\left(2.755\times 10^{2}\right) \left(\frac{m_\chi}{\text{keV}}\right) Y_\chi^{\rm today} \propto m_\chi \frac{\epsilon^{4n-2}~n^2}{M_S}   \propto v_{S} \frac{\epsilon^{6n-3}~n^2}{M_S}  .
\end{eqnarray}
Therefore for fixed values of $M_S$ and $n$, $Y_\chi^{\rm today}$ becomes constant, and the relic density increases with the increase of the DM mass $m_\chi$ as $\Omega_\chi h^2 \propto m_\chi$ or $\Omega_\chi h^2 \propto v_S$.  As depicted in the left panel of Fig. \ref{fig:6}, with the increase of $M_S$ and $n$, $Y_\chi$ decreases resulting in the decrease of  DM relic density with fixed $v_S$  as $\Omega_\chi h^2 \propto \epsilon^{6n-3}/M_S$. These features are also portrayed in the right panel of Fig. \ref{fig:6}. The observed relic density constraint measured by PLANCK is shown by the black dashed line in the same plane. 
%The gray-shaded region with $m_{\rm DM} < 5.3$ keV is excluded by the structure formation constraint(Lyman-$\alpha$ line).
\begin{figure}
    \centering
    $$
\includegraphics[width=7cm]{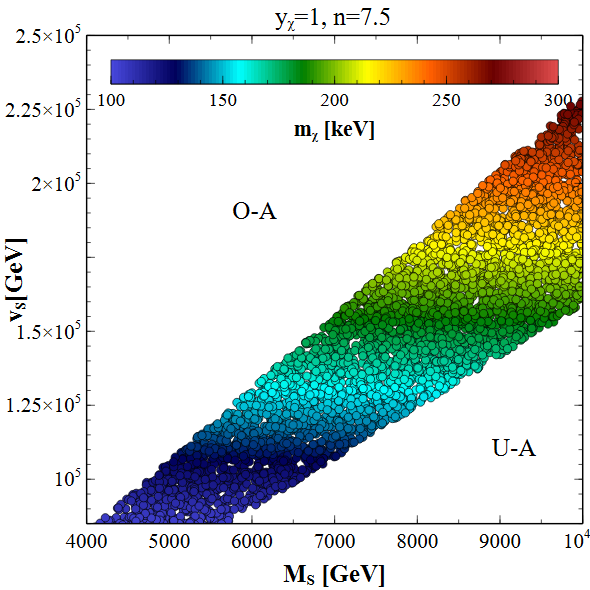}~~
\includegraphics[width=7cm]{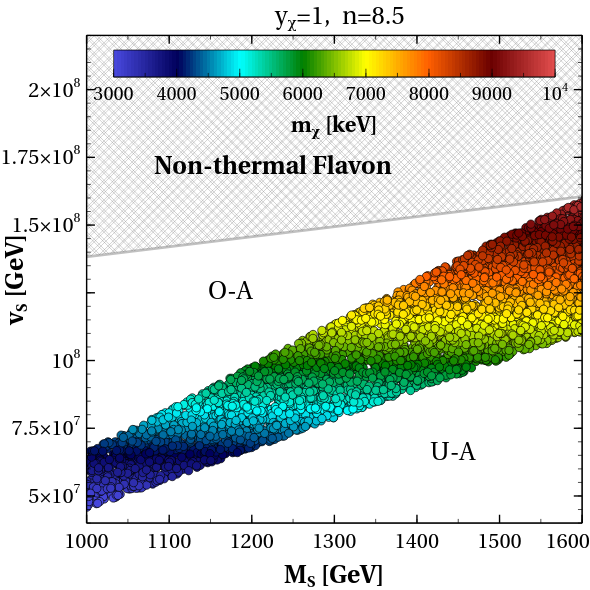}
    $$
    \caption{ \it Observed dark matter abundance parameter space is shown in  $v_{S}-M_{S}$ plane for two different choices of $n$ values: (a) $n=7.5$ (left panel) and (b) $n=8.5$ (right panel). The corresponding $m_{\chi}$ values are shown in the color bar. The region above and below the colored region corresponds to over- (O-A) and 
    under-abundant (U-A) regions respectively.}
    \label{fig:7}
\end{figure}

In Fig. \ref{fig:7} we show the observed relic density satisfying in the plane of $M_{S}-v_{S}$ for $n=7.5$ (left panel) and $n=8.5$ (right panel). For $n=8.5$, the Yukawa coupling, $\epsilon^{2n-1}$, is more suppressed compared with $n=7.5$. Therefore for a fixed $v_S$ (or fixed DM mass $m_\chi$), $M_S$ has to decrease with an increase in $n$ in order to obtain the correct DM relic density.  The correct relic density constraint regulates the DM and flavon masses for our chosen values of $n$.  
%================================
\section{Pseudoscalar Flavon Portal for Dark Matter Freeze-in}
\label{sec:pseudo}

So far we focused on a scenario for DM freeze-in where both the scalar and the pseudosclar components of the flavon field $S$ played an equal and  essential role. In fact, we treated the two states to be degnerate in our numerical studies. Such a scenario requires the  VEV of $S$ to be much larger than the mass of $h_S$ and $A_S$, which in turn requires  $\lambda_S  \sim (m_{h_S}^2/v_S^2) \ll 1$
on the self-coupling of $S$, see Eq. (\ref{eq:param}).  In this section we investigate a scenario where $\lambda_S\sim {\cal O}(1)$ is chosen. This would imply that $v_S$ is of the same order as $m_{h_S}$. In this case the pseudoscalar component $A_S$ can serve as the DM freeze-in portal.  Note that the mass of $A_S$ arises solely from the soft $U(1)$ breaking parameter $\mu_b^2$ (cf. Eq. (\ref{eq:param})).  Suppose that the reheat temperature after inflation $T_R$ is below the mass of $h_S$.  In this case $h_S$ interactions won't bring the DM field into thermal equilibrium via freeze-in, since $h_S$ is never produced post-inflation and is not in the thermal bath.  The pseudoscalar field $A_S$, with its mass assumed to be of order TeV, is in the thermal bath, and in equlibrium with the SM particles via its Yukawa interactions.  As for any pseudo-Goldstone bosons, it is convenient to express the flavon field in an exponential parametrization
\begin{equation}
S = \left(\frac{h_S + v_S}{\sqrt{2}}\right) \,e^{i A_S/v_S},
\end{equation}
which upon expansion of the soft breaking term in $V(H,S)$ of Eq. (\ref{eq:pot}) leads to the scalar interactions of $A_S$ given as
\begin{equation}
    V(H,S) = \frac{M_{A_S}^2}{4}(h_S+v_S)^2\cos\left( \frac{2 A_S}{v_S}\right) + ...
\end{equation}
When the cosine function is expanded, one sees that the self-interactions of $A_S$ are suppressed by $(M_{A_S}/v_S)^2 \ll 1$, as well as the interactions of $A_S$ with $h_S$.  Although $h_S$ could keep the DM $\chi$ in thermal equilibrium with the SM plasma via the $\lambda_{HS}$ coupling of Eq. (\ref{eq:pot}), assuming that this coupling is of order one, in the scenario we propose here $h_S$ has a mass above $T_R$, and thus cannot bring $\chi$ into thermal equilibrium. Note that $A_S$ can be thermalized at high temperature in a way similar to the one in the previous section. Therefore DM can only be produced from the light thermal pseudoscalar $A_S$ similarly to the previous case. 

There is an independent reason to take the reheat temperature $T_R$ to be below $v_S$.  If $T_R$ were above $v_S$, then the Froggatt-Nielsen fermion fields that are responsible for inducing the effective Lagrangian of Eqs, (\ref{eq:FN}) and (\ref{LDM}) would be brought into thermal equilibrium with the plasma.  This would in turn bring the DM $\chi$ also into thermal equilibrium, which is contrary to the FIMP hypothesis.

\begin{figure}
$$
\includegraphics[scale=0.5]{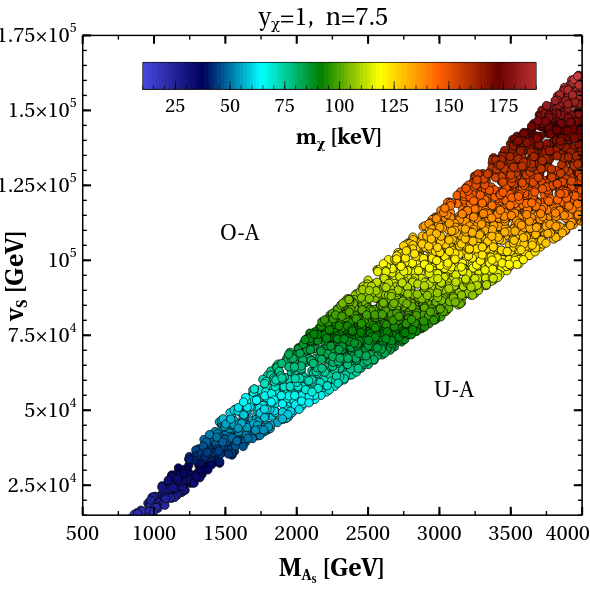}~~
\includegraphics[scale=0.5]{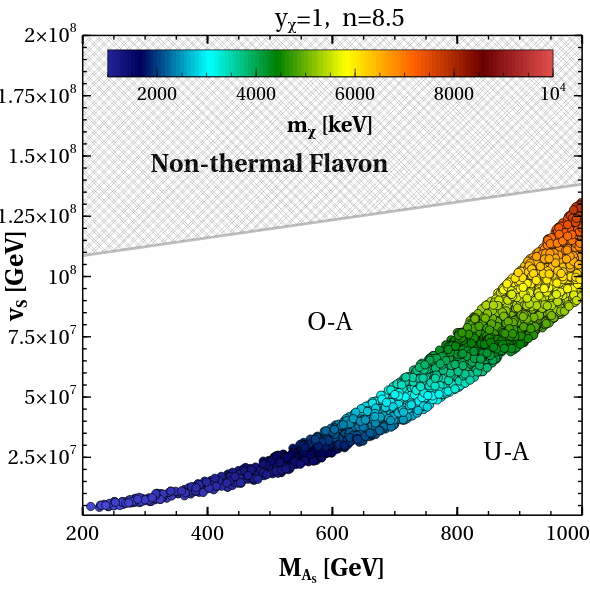}
$$
 \caption{\it Parameter space satisfying dark matter relic density in  $v_{S}-M_{A_{S}}$ plane for two different choices of $n$ values: (a) $n=7.5$ (left panel) and (b) $n=8.5$ (right panel). The corresponding $m_{\chi}$ values are shown in the color bar. The regions above and below the colored region correspond to over- (O-A) and under-abundant (U-A) regions respectively.}
 \label{fig:pseudoprod}
\end{figure}

We shall now briefly discuss the allowed region of DM parameter space in the context of pseudoscalar flavon portal. In Fig.~\ref{fig:pseudoprod}, we show the production of DM through pseudo-scalar $A_S$ keeping $T_R 
< v_S \simeq m_{h_S}$. Here we show the region of parameters satisfying the observed relic density  in the $M_{A_S}- v_S $ plane keeping $n=7.5$ in left panel and $n=8.5$ in the right panel. The color bar shows the corresponding value of $m_{\chi}$ for every $v_S$. The region under (over) the colored area corresponds to the under-abundance (over-abundance) of dark matter density.  We see a behavior similar to the results of Fig. \ref{fig:7}.

%===========================
\section{Discussion on free streaming length}
\label{sec:FSL}
Understanding the behavior of dark matter's free streaming is of utmost importance as it plays a critical role in the formation of structures. A larger free streaming length reduces the likelihood of structure formation on scales similar to that length. The free streaming length is inherently connected to the average momentum of dark matter particles and can be estimated by integrating from the time of decoupling to the present time.

When dark matter particles possess substantial initial momentum, their free streaming effect can eliminate structures on scales smaller than the free streaming horizon, denoted as $\lambda_{{\rm FSL}}$. In the present framework, the dark matter mass ranges from sub-MeV to sub-GeV, depending on the choices of the FN charge $n=7.5$ and $n=8.5$. Additionally, the dark matter is generated from a TeV-order flavon ($S$). In such scenarios, certain constraints on the free streaming length ($\lambda_{{\rm FSL}}$) for freeze-in dark matter may be applicable based on the requirements of structure formation. While hot dark matter has already been ruled out, warm dark matter with a free streaming length $\lambda_{{\rm FSL}} < 0.1~{\rm Mpc}$ remains acceptable~\cite{Drewes:2016upu}. The phenomenon of free streaming length associated with the production mechanism of dark matter has been extensively investigated in various studies \cite{Ghosh:2022hen,BhupalDev:2013oiy,Biswas:2022vkq}.

\iffalse
As in our case, the dark matter is light(keV/MeV) and is produced from a very heavy particle(TeV), it may face some constraint from free streaming length. 
Though "hot" dark matter is excluded, warm dark matter  with free streaming length $\lambda_{\rm FSL}<0.1 Mpc$ is still allowed. 
This aspect of free streaming length depending on the production mechanism of DM has been discussed in several works \cite{Ghosh:2022hen,Biswas:2022vkq}. 
\fi

Following the analysis of ref \cite{Ghosh:2022hen}, we make an educated guess of the co-moving free streaming length of the DM particle, which refers to the distance covered by dark matter (DM) particle in co-moving coordinates from the scale of decoupling of scattering reactions $(a_{\rm dec})$ to the scale of matter-radiation equality $(a_{\rm eq})$: 
\begin{equation}
    \lambda_{\rm FSL} \simeq 0.0394 {\rm Mpc} \int_{a_{dec}}^{a_{eq}}  ~da~ \frac{\langle v(a)\rangle}{a^{\frac{3}{2}}_{eq} ~\sqrt{a+a_{eq}}} ~,
\end{equation}
where $\langle v(a) \rangle$ is the average velocity defined as
\begin{equation}
    \langle v(a) \rangle = \int dq \frac{q^3 f_{\chi}(q,T)}{\sqrt{q^2+ \frac{m^2_{\chi} a^2}{T^2_0}(\frac{g_*(T_{dec})}{g_*(T_0)})^{2/3}}}/\int dq q^2 f_{\chi}(q,T)
\end{equation}
Here $q$ is a scaled variable which includes the effect of red-shift of the momentum variable $p$ through the relation $q= a \frac{p}{T_0}(\frac{g_*(T_{dec})}{g_*(T_0)})^{2/3}$. $T_{dec}$ in our case is decided by the mass of the flavon, $T_0$ is current temperature of the universe and $a_{eq}$ is taken to be $2.91 \times 10^{-4}$. Following the ref. \cite{Ghosh:2022hen}, we evaluate the momentum distribution function $f_{\chi}(q,T)$ followed by the estimation of average DM velocity. Eventually, we get $\lambda_{\rm FSL}$ by substituting $\langle v(a)\rangle $ in
Eq.(5.1) and doing the integration numerically for different set of parameters as shown in Table~\ref{tab:my_label}.
\begin{table}[ht]
    \centering
    \caption{Free streaming length at relic satisfied parameter points}
    \begin{tabular}{p{3cm} p{3cm} p{3cm} p{3cm}  }
\hline
\multicolumn{4}{c}{Case I: DM from Degenerate CP even and odd states} \\
\hline \hline
n & $M_{S}$(TeV) & $m_{\chi}$(keV) & $\lambda_{\rm FSL}$(Mpc) \\
 \hline
  7.5  & 5.0 & 125  & 4.66 \\
 8.5 & 1.0 &  $3\times 10^3$ &  0.06 \\
8.5  & 1.5 &   $9 \times 10^3$  &  0.02 \\
\hline\hline
\multicolumn{4}{c}{Case II: DM only from the pseudoscalar portal} \\
\hline \hline
n & $M_{A_S}$(TeV) & $m_{\chi}$(keV) & $\lambda_{\rm FSL}$(Mpc) \\
\hline
 7.5 & 1.0 & 25 &  3.97 \\
 8.5 & 0.3  & $2 \times 10^3$ & 0.01 \\
8.5 & 1.0  & $8\times 10^3$ & 0.02 \\
\hline
 \end{tabular}
    \label{tab:my_label}
\end{table}
The first three rows of the table correspond to the case where both CP even and CP odd states of flavon are considered whereas the last three rows correspond to the case where the DM is produced from pseudoscalar only. 
%The effect of the decay term $(S \to \chi \chi)$ is only taken into account 
 %while calculating the free streaming length of DM. 
 From this table it is clear that $\lambda_{\rm FSL}$ mostly depends 
 on the dark matter mass and it decreases with the increase in the dark matter mass for a given value of $M_S(M_{A_S})$. Heavier
DM particle becomes non-relativistic much earlier consequently leading to smaller free streaming length. From the aforementioned calculation of $\lambda_{\rm FSL}$ one can see that the DM particle with mass up to order of a few hundred keV can be disfavoured from the constraint 
on the free streaming length $\lambda_{\rm FSL} < 0.1 $ Mpc. 
%with mass up to By imposing the bound for warm dark matter, we can %rule out the scenario $n=7.5$ where the dark matter mass is in keV %ranges though the $n=8.5$ scenario remains still allowed as the masses %in MeV ranges does not affect the structure formation. 
It is important to note that in estimating the 
 $\lambda_{\rm FSL}$ of DM we only consider the production of DM from the decay of heavy flavons $(S \to \chi \chi)$ for simplicity. However, it is worth mentioning that DM $(\chi)$ can also be produced from the scattering of heavy flavons $SS \to \chi \chi$ and ${\rm SM~SM} \to \chi \chi$. Inclusion of  these processes in the calculation of $\lambda_{\rm FSL}$ can potentially reduce the value of co-moving free streaming length approximately by a factor of $\sim 40$ as shown in ref.\cite{Ghosh:2022hen}, thus marginally allowing keV range 
 DM. Hence, the values of $\lambda_{\rm FSL}$ shown in Table \ref{tab:my_label} should be
considered indicative. A detailed and precise calculation of the co-moving free streaming length of the DM within this scenario is beyond the scope of this analysis and will be reported elsewhere \cite{Babu}. 
 
%===========================
\section{Conclusion}
\label{sec:concl}
%====================
In this work, we have proposed a unified solution to the fermion mass hierarchy and a FIMP dark matter within a class of $U(1)_{\rm FN}$ extensions of the Standard Model.  The DM candidate $\chi$, assumed to be a Majorana fermion in this work, carries a $U(1)_{\rm FN}$ charge, and acquires a suppressed mass by the same dynamics as the FN mechanism.  The flavon field $S$ connects $\chi$ with the SM sector, but the Yukawa coupling of $S$ with $\chi$ is suppressed by a small parameter $\epsilon^{2n-1}$ where $n$ is the $U(1)_{\rm FN}$ charge of $\chi$. We have presented two scenarios for the $U(1)_{\rm FN}$ charges, and showed how the dark matter fermion $\chi$ could be incorporated in both these two
 scenarios. Our model parameters satisfied constraints coming from vacuum stability, perturbative unitarity, invisible Higgs decay, exotic heavy Higgs search and heavy flavon induced flavor violating processes. In addition to these, our model parameters are also restricted by constraints coming from dark matter relic density. After taking into account all those constraints, our scenario naturally leads to a FIMP dark matter, corresponding to relatively large values of $n = 6.5-8.5$. 
In this case it is natural for the DM to be relatively light, due to the suppression by the FN mechanism for its mass. The DM never achieved thermal equilibrium in the early universe and is produced from the decays of some heavy fields present in the thermal bath. In this work we have explored the viable parameter space for the freeze-in DM production from the decays of the flavon $S$ responsible for generating mass hierarchies of SM fermions. We have shown a preferred range for the DM mass, which is  $(100-300)$ keV and $(3-10)$ MeV, corresponding to $n = 7.5$ and $8.5$ respectively. 
We have also shown that by choosing the charge $n$ of $\chi$ to be a half-integer, while the rest of the fields have integer charges, the DM becomes stable, thus explaining an important question for any DM model. 
 We have also computed the co-moving free streaming length of the DM in this scenario and shown that DM mass upto few hundreds of keV is disfavoured from limit of the free streaming length.

\section*{Acknowledgements}
ND, DKG and PG 
would like to thank Prof. Satyanarayan Mukhopadhyay and Deep Ghosh for useful discussion. SC would like to thank IACS for local hospitality where this work was initiated. The work of SC is supported by the College of Holy Cross Bachelor Ford Summer fellowship '21-'22.
The work of KSB is supported in part  by the U.S. Department of Energy  under grant number DE-SC0016013. ND is funded by CSIR, Government of India, under the NET SRF
fellowship scheme with Award file No.09/080(1187)/2021-EMR-I.

%===========================================================
%\clearpage
%%%%%%%%%%%%%%%%%%%%%%%
\appendix
\section{Appendix}
%=====================
\subsection{Interactions}
\label{apx:int}
%======================
%==========================
In this section we collect the couplings of flavon (S) with  DM $\chi$ .  These are given by:
\begin{eqnarray}
C_{S \chi \chi} &=&  \sqrt{2} \,n \, y_{\chi}\epsilon^{2n-1} \nonumber \\
C_{h_S \chi \chi} &=&  \sqrt{2} \,n \, y_{\chi}\epsilon^{2n-1} \nonumber \\
C_{A_S \chi \chi} &=&  \sqrt{2} \,n \, y_{\chi}\epsilon^{2n-1}\nonumber \\
C_{S S \chi \chi } &=& \frac{y_{\chi} n (2n-1)\epsilon^{2n-1}}{\sqrt{2} v_S}~.
\end{eqnarray}
%=========================
%=========================
\subsection{Decay widths of scalars}
\label{apx:dw}
%======================
%==========================
%Here we present the decay widths of the two scalars $h_1$ and $h_2$ into various channels. In the electroweak symmetric vacuum we use the decay width of $h_S$, rather than those of $h_{1.2}$. The decays into $\overline{\chi^c} \chi$ serves as one of the production mechanisms for the DM fermion.

The decay width of the flavon into DM is given by
\begin{eqnarray}
    \Gamma(S\to \chi \chi)&=&\frac{M_S}{16\pi}\Big(1-\frac{4 m^2_\chi}{M^2_{S}}\Big)^\frac{3}{2} ~{(C_{S \chi \chi})}^2 ~.
\end{eqnarray}

%%%%%%%%%%%%%%%%%%
\subsection{Interaction rates}
\label{apx:IntRate}
%======================
%==========================
Here we have gathered the expressions for all the interaction rates  as mentioned in the text. The interaction rates of the flavon as well as the DM is of our main interest. First, we discuss the interaction rates of flavon. Interaction rates arising from the Dim 6 Yukawa interaction and the Higgs portal interaction  are given by
\begin{eqnarray}
 {\bf \Gamma}_{SS \leftrightarrow q \Bar{q} H} &=& n^{eq}_{S}~ \langle\sigma v \rangle_{SS \to q \Bar{q} H} \nonumber \\
{\bf \Gamma}_{SSH \leftrightarrow q \Large{\bar{q}}} &=& n^{eq}_{S}~n^{eq}_{H}~ \langle\sigma v^2 \rangle_{SSH \to q \bar{q}} \nonumber \\
{\bf \Gamma}_{S H \leftrightarrow q \Bar{q}}&=& n^{eq}_{H}~ \langle\sigma v\rangle_{SH \to q \Bar{q}}\nonumber \\
{\bf \Gamma}_{S q \leftrightarrow q H}&=& n^{eq}_{q}~ \langle\sigma v\rangle_{S q \to q H}\nonumber \\
{\bf \Gamma}_{S S \leftrightarrow H H}&=& n^{eq}_{S}~ \langle\sigma v\rangle_{SS \to H H} \nonumber \\
{\bf \Gamma}_{S S \leftrightarrow \chi \chi}&=& n^{eq}_{S}~ \langle\sigma v\rangle_{SS \to \chi \chi}
\end{eqnarray}
%====
\color{black}
where $n^{eq}_x$ is the equilibrium density of the particle x. The thermally averaged cross-section of the $2 \to 2$ processes i.e $SS \rightarrow HH$,$SS \rightarrow \chi \chi$,$ SH \rightarrow q \Bar{q}$ and $ Sq \rightarrow qH$ are mentioned below. 
\color{black}
\begin{equation}
\langle\sigma v\rangle_{SS \to H H} =\frac{1}{8\pi M^4_S\, T K^2_2(\frac{M_{S}}{T})} \int_{4 M^2_S}^{\infty} \sigma(s) (s-4 M^2_S)\sqrt{s}\, K_1(\frac{\sqrt{s}}{T})\, ds 
\end{equation}
where the cross-section of the process is given by
\begin{equation}
  \sigma(s) =  \frac{\lambda^2_{HS}}{16 \pi s}\, \frac{\sqrt{s- 4 m^2_H}}{\sqrt{s- 4 M^2_S}} ~.
\end{equation}
The thermal average of $S H \to q \bar{q}$ process is given by
\begin{equation}
    \langle\sigma v\rangle_{S H \to q \Bar{q}} =\frac{1}{8\pi M^2_S m^2_H\, T K_2(\frac{M_{S}}{T}) K_2(\frac{m_H}{T})} \int_{
{s^2_0}}^{\infty} \sigma_{SH \to q \bar{q}} (s-{s^2_0}) \sqrt{s}\, K_1(\frac{\sqrt{s}}{T})\, ds
\end{equation}
where the lower limit of the integral is $s_0= M_{S}+m_{H}$ and the cross-section of the above process is given by 
\begin{eqnarray*}
 \sigma_{SH \to q \bar{q}} &=& \frac{C^2_{SHqq}}{\pi} \frac{s}{\sqrt{(s-{(M_S+m_H)}^2)(s-{(M_S-m_H)}^2)}} ~.
\end{eqnarray*}
The thermally averaged cross section of $S q \to q H$ is given by
\begin{equation}
    \langle\sigma v\rangle_{S q \to q H} =\frac{1}{8\pi M^2_S m^2_q\, T K_2(\frac{M_{S}}{T}) K_2(\frac{m_{q}}{T})} \int_{s^2_{1}}^{\infty} \sigma_{Sq \to q H} (s-s^2_{1}) \sqrt{s}\, K_1(\frac{\sqrt{s}}{T})\, ds
\end{equation}
where $s_{1}=(M_S+m_q)$ and the cross section is given by
\begin{equation}
  \sigma_{Sq \to q H}=  \frac{C^2_{SHqq} }{8 \pi } \frac{(s-m^2_H)^2}{s^2}~.
\end{equation}
Here $C_{SHq\Bar{q}}$ is the coupling of the processes. From Eq. (\ref{eq:int2}), we can see that there can be different $2 \to 2$ number changing processes involving $SH \to \tau \bar{\tau}$,$SH \to b \bar{b}$,$SH \to c\Bar{t}$,$S \tau \to \tau{H}$,$S b\to b H$, $S c\to t H$ and $S t\to c H$. We have considered the exact values of masses and coupling  mentioned in the paragraph followed by Eq. (\ref{eq:int2}). Though  $a^{(u)}_{23}$ and $a^{(u)}_{32}$ does not have determined values, we have taken them to be 1.
The thermally averaged cross section of  and $SS \rightarrow q \Bar{q} H$ ($2 \rightarrow 3$), $ SSH \rightarrow q \Bar{q}$($3 \rightarrow 2$) are given by \cite{Barman:2020plp}
\begin{eqnarray}
\langle \sigma v \rangle_{SS \leftrightarrow q \Bar{q} H}  &=&\frac{1}{\prod_{i=1}^{2} n^{eq}_{i}} \int 
\prod_{i=1}^{5}d \Pi_i (2 \pi )^4 \delta^4(p_1+p_2-p_3-p_4-p_5)\Bar{|M|^2} \prod_{i=1}^3 f_i \nonumber \\
&=& \frac{1}{\prod_{i=1}^{2} n^{eq}_{i}} \frac{T}{64 \pi^4} \int_{4 M_{S}^2}^{\infty} \sqrt{s} K_1 (\frac{\sqrt{s}}{T}) ds \int_{2 m^2_q}^{(\sqrt{s}-m_{H})^2} \frac{ds_{23}}{2 \pi} \frac{8 s_{23}}{\Lambda^4}\frac{\Bar{\beta}(y_1,y_2)}{8 \pi} \frac{\Bar{\beta}(y_3,y_3)}{8 \pi}\nonumber \\
\label{eqn:2 to3}
\end{eqnarray}
where $\Bar{\beta}$ is defined by
\begin{equation*}
    \Bar{\beta}(x_1,x_2)=\sqrt{1+x^2_1+x^2_2-2 x_1 -2 x_2- 2 x_1 x_2}
\end{equation*}
and 
\begin{eqnarray*}
\Bar{\beta}(y_1,y_2) &=& \Bar{\beta}(\frac{m^2_{H}}{s},\frac{s_{23}}{s}) \\
\Bar{\beta}(y_3,y_3) &=&  \Bar{\beta}  (\frac{m^2_{q}}{s_{23}},\frac{m^2_{q}}{s_{23}})
\end{eqnarray*}
where $8 s_{23}/\Lambda^4$ is the squared  amplitude of the process and the rest of the integrand comes from the phase spaces and energy distribution functions. 
\begin{eqnarray}
\langle \sigma v^2 \rangle _{SSH \rightarrow q \Bar{q}} &=& \frac{1}{\prod_{i=1}^{3} n^{eq}_{i}} \int \prod_{i=1}^{5}d \Pi_i  (2 \pi )^4 \delta^4(p_1+p_2+p_3-p_4-p_5)\Bar{|M|^2} \prod_{i=1}^3 f_i \nonumber \\
&=& \frac{1}{\prod_{i=1}^{3} n^{eq}_{i}} \frac{T}{64 \pi^4} \int \sqrt{s} K_1 (\frac{\sqrt{s}}{T}) \int_{2 m^2_q}^{(\sqrt{s}-m_{H})^2} \frac{ds_{23}}{2 \pi} \frac{8 s}{\Lambda^4}\frac{\Bar{\beta}(y_2,y_3)}{8 \pi} \frac{\Bar{\beta}(y_1,y_1)}{8 \pi} \nonumber \\
\end{eqnarray}
where 
\begin{eqnarray*}
\Bar{\beta}(y_1,y_1)&=& \Bar{\beta}(\frac{M^2_{S}}{s_{23}},\frac{M^2_{S}}{s_{23}}) \\
\Bar{\beta}(y_2,y_3) &=& \Bar{\beta}(\frac{m^2_{H}}{s},\frac{s_{23}}{s})~.
\end{eqnarray*}
Next we discuss about the interaction rates of our DM candidate,$\chi$. The interaction rate for $S \to \chi \chi$ process is defined by
\begin{equation}
    \langle \Gamma_{S \rightarrow \chi \chi} \rangle = \frac{\int \Gamma e^{-\frac{E_1}{T}} d^3 p_1}{\int e^{-\frac{E_1}{T}} d^3 p_1} = \frac{I_1}{I_2}
\end{equation}
 where $\Gamma_{S \to \chi \chi}$ is the rest frame decay width of the process.
 The numerator of this expression is given by
\begin{eqnarray*}
   I_1  &=& \int d^3 p_1 \, e^{{-\frac{E_1}{M_S}. \frac{M_S}{T}}} \frac{1}{ 2 E_1} \int d \Pi_1 d \Pi_2  (2 \pi )^4 \delta^4(p_1-p_2-p_3)\overline{|M|^2} \nonumber \\
   &=& 4 \pi M^3_S \int \sqrt{x^2-1} \,~ e^{-x. \frac{M_S}{T}} dx\,\frac{1}{ 2 M_S} \int d \Pi_1 d \Pi_2  (2 \pi )^4 \delta^4(p_1-p_2-p_3)\overline{|M|^2} \nonumber \\
   &=& 4 \pi M^3_S \frac{K_1(M_S/T)}{(M_S/T)}\,\Gamma_{S\to \chi \chi}=4 \pi M^2_S \,T\, K_1(M_S/T)\,\Gamma_{S\to \chi \chi} \nonumber
\end{eqnarray*}
where $x=\frac{E_1}{M_S}$.
The denominator of the expression is given by
 \begin{eqnarray}
   I_2= &=& \int e^{-\frac{E_1}{T}} 4 \pi p_1 E_1 dE_1 =4 \pi M^3_S \int  e^{-x.\frac{M_S}{T}}\sqrt{x^2-1} \,x \, dx \nonumber \\
   &=& 4 \pi M^3_S \,\frac{K_2(M_S/T)}{(M_S/T)}=
   4 \pi M^2_S T K_2(M_S/T)~.
 \end{eqnarray}
Therefore the averaged decay width of $S \to \chi \chi$ is given by
\begin{eqnarray}
     \langle \Gamma_{S \rightarrow \chi \chi} \rangle &=& \frac{K_1(M_S/T)}{K_2(M_S/T)} \,\Gamma_{S\to \chi \chi}~.
\end{eqnarray}
The thermal average of $SS \to \chi \chi$ process is given by
\begin{equation}
    \langle \sigma v \rangle_{SS \to \chi \chi} =\frac{1}{8\pi M^2_S M^2_S\, T K_2(\frac{M_{S}}{T}) K_2(\frac{M_S}{T})} \int_{
 4 {M_S}^2}^{\infty} \sigma_{S S \to \chi \chi} (s-4 {M_S}^2) \sqrt{s}\, K_1(\frac{\sqrt{s}}{T})\, ds
\end{equation}
where the cross-section is given by
\begin{equation}
    \sigma_{S S \to \chi \chi}= \frac{C^2_{SS\chi \chi} }{16 \pi} \frac{\sqrt{s- 4 m^2_{\chi}}}{\sqrt{s-4 M^2_S}}~.
\end{equation}
%%%%%%%%%%%%%%%%%%%%
\bibliographystyle{JHEP}
\bibliography{Bibliography}
%%%%%%%%%%%%%%%%%%
\end{document}